\documentclass[]{aa}
\usepackage[varg]{txfonts}
\usepackage{graphicx}
\RequirePackage[colorlinks=true,linkcolor=blue,citecolor=blue,urlcolor=blue]{hyperref}
\newcommand{\dcdot}{\mathbin{%
    \nonscript\mspace{-\muexpr\medmuskip*2/3}%
    \cdot
    \nonscript\mspace{-\muexpr\medmuskip*2/3}%
  }%
}

\begin{document}

\title{IllustrisTNG + Cosmic Rays with a Simple Transport Model:\\ From Dwarfs to L$^\star$ Galaxies}

\author{Rahul Ramesh\thanks{E-mail: rahul.ramesh@stud.uni-heidelberg.de}
\and Dylan Nelson
\and Philipp Girichidis} 
\institute{
Universität Heidelberg, Zentrum für Astronomie, ITA, Albert-Ueberle-Str. 2, 69120 Heidelberg, Germany \label{1}
}

\date{}

\abstract{We use a simple model for cosmic ray (CR) production and transport to assess the impact of CRs on $z$\,$=$\,$0$ galaxy, circumgalactic medium (CGM), and halo properties. To do so, we run the first suite of large-volume cosmological magnetohydrodynamical simulations (25\,Mpc\,h$^{-1}$ boxes) with the IllustrisTNG galaxy formation model including CR physics. We select CR transport parameters that yield a reasonable trade off between realistic large-scale integrated properties, and galactic CR pressure profiles predicted by more complex models. The resulting simulations show that, at fixed halo mass, including CRs does not strongly impact the temperature, density, or (total) pressure structure of the CGM with respect to the fiducial TNG model. However, cosmic rays add significant non-thermal pressure support to the halo. This suppresses the star formation activity and thus stellar masses of galaxies, from dwarf to L$^\star$ halos. The cosmic star formation rate density, stellar mass function, and stellar mass to halo mass relation are all reshaped by CRs. Galaxy sizes and halo-scale gas fractions are more mildly affected, while lower gas densities in the interstellar medium inhibit supermassive black hole growth. Halo gas is also less magnetized, and less metal enriched. These differences are primarily driven by suppressed gas accretion onto halos and galaxies, as well as weaker galactic outflows in the presence of CRs. Our findings are in qualitative agreement with previous studies of the impact of CRs on galactic outflows, and motivate the inclusion of CR physics in future large-scale cosmological simulations.}


\titlerunning{IllustrisTNG + Cosmic Rays: A Simple Transport Model}
\authorrunning{R. Ramesh, D. Nelson \& P. Girichidis}

\maketitle


\section{Introduction}

Cosmological simulations have become powerful tools for understanding the physics of cosmic structure formation. Some of these numerical calculations include only gravity, thereby simulating the evolution of mock universes comprised solely of (collisionless) dark matter particles. However, recent simulations now commonly include baryons and many of the associated physical processes that are considered important for galaxy formation and evolution \citep[for a recent overview of cosmological simulations, see][]{vogelsberger2020}.

For instance, it is common practice to include models for the radiative cooling of gas, either with a simple primordial network \citep[e.g.][]{katz1996} or incorporating additional processes such as metal line cooling \citep[e.g.][]{wiersma2009}. This can be paired with a model for a time-dependant UV background \citep[e.g][]{fg2009,haardt2012} with corrections for self-shielding \citep[e.g.][]{rahmati2013}. As gas reaches sufficiently high densities, star formation can occur, creating stellar particles that can represent individual stars \citep[e.g.][]{gutcke2021,smith2021}, or entire populations \citep[e.g.][]{springel2003}.

Galactic feedback processes are an important mechanism to regulate the rate of star formation (SFR) and the amount of stellar mass growth. In low-mass galaxies (M$_\star$\,$\lesssim$\,$10^{10.5}$\,M$_\odot$), outflows driven by stellar and supernovae feedback play an important role \citep[e.g.][]{fielding2017}. In addition to keeping the SFR in check, these also inject significant amounts of energy and metals into the surrounding halo gas \citep[e.g.][]{nelson2019,pandya2021}. Higher mass galaxies are instead dominated by the activity of the central supermassive black hole (SMBH), which may release energy in thermal, kinetic, or other physical channels \citep[e.g.][]{schaye2015,weinberger2017,dave2019}. Such feedback episodes launch gas out of galaxies at high velocities \citep[e.g.][]{oppenheimer2020,ramesh2023a}, alter the morphology and kinematics of the circumgalactic medium \citep[CGM;][]{nelson2015,kauffmann2019}, increase the cooling time of gas in the surrounding halo \citep{zinger2020} and hinder future gas accretion and star formation \citep[e.g.][]{davies2020}.

In addition, recent simulations now start to include important physics arising from non-thermal processes. Most notably, magnetic fields \citep[e.g.][]{pakmor2017,marinacci2018} and their impact on the propagation of galactic outflows \citep{steinwandel2020} as well as halo-scale gas properties \citep{pakmor2020}. In addition, theory suggests that magnetic fields may play an important role in the evolution of small-scale CGM clouds: either by suppressing fluid instabilities \citep{cottle2020,sparre2020,ramesh2024c}, providing additional pressure support \citep{nelson2020,ramesh2023b,fielding2023} or, in the presence of a draped layer \citep{dursi2008,pfrommer2010}, boosting the drag force they experience \citep{mccourt2015,ramesh2024b}. Such cool clouds may play a part in replenishing the galactic cold gas supply \citep{lepine1994}, making their survival and evolution unavoidably linked to the star-formation activity of galaxies.

The second non-thermal physical component of significant interest for galaxy evolution is cosmic rays (CRs). The energy density of CRs in galaxies is non-negligible, and in rough equipartition with magnetic, thermal and turbulent energies \citep{cox2005,naab2017}. Depending on their momenta, CRs can influence gas in various ways. Cosmic ray particles in the $\lesssim$\,$100$\,MeV\,$c^{-1}$ range of the spectrum can ionize their surrounding gas \citep[e.g.][]{ivlev2018,padovani2020}, while CRs with $\sim$\,GeV\,$c^{-1}$ can drive powerful galactic-scale outflows \citep[e.g.][]{zirakashvili1996,girichidis2016,chan2022} through gradients in their pressure structure \citep{pfrommer2017,butsky2020}. Much like magnetic fields, this added non-thermal pressure support may also provide additional stability to small-scale cold gas structures \citep{bruggen2020,butsky2022}.

More energetic CRs ($\gtrsim$\,TeV\,c$^{-1}$) are not dynamically relevant, as they are sub-dominant in terms of their contribution to the total energy budget. However, pion production via hadronic interactions leads to a subsequent decay into $\gamma$-rays, providing an important and observable non-thermal radiative signature \citep{kotera2011,werhahn2021a,werhahn2021b}. While these high energy CRs are less impacted by streaming losses \citep{girichidis2024}, their lower energy counterparts excite Alfv\'en-waves through the CR streaming instability when they propagate faster than the Alfv\'en-velocity \citep{kulsrud1969}, effectively heating up gas as these waves damp on short-time scales \citep{wiener2017,ruszkowski2017,thomas2023}.

CRs may thus play an important role in the evolution of galaxies and their gaseous halos. They may have an impact on the accretion of gas onto galaxies by altering the flow structure of CGM gas, thereby modifying structural properties such as galactic disk sizes \citep{buck2020}, and also time-integrated quantities like the star formation rates within the ISM \citep{farcy2022}. Outflows driven by CRs can be smoother, cooler and denser than those by supernovae \citep{girichidis2018, armillotta2024}, which may in turn affect the escape of Ly$\alpha$ photons from galaxies \citep{gronke2018}. Their added non-thermal pressure may result in halos that are significantly cooler \citep{ji2020}. A number of studies have also suggested that they may play a role in magnetizing gas through the so called galactic dynamo \citep{parker1992,hanasz2009,pfrommer2022}.

Thus far, the impact of CRs has been studied in isolated (idealized) galaxy setups, and in cosmological zoom-in simulations of single halos. However, their impact on a cosmologically representative galaxy population remains unknown: they have never been studied in a large-volume cosmological (magneto)hydrodynamical simulation. This is partly due to the small timesteps required to track their evolution \citep{hopkins2017,chan2019}, as well as the complexity involved in representing CRs over a broad spectral distribution \citep[e.g.][]{girichidis2020,hopkins2023a}, making it challenging to include this component in already computationally expensive simulations \citep[see also][]{jiang2018,gupta2021}.

To overcome this limitation, \cite{hopkins2023b} recently proposed a sub-grid `toy' recipe aimed at capturing the leading first-order impact of CRs on galaxy growth. The model is parameterized by two simple quantities, and the implementation is similar to the gravity-tree algorithm \citep{barnes1986}, thereby allowing one to estimate the CR contribution at any point in space with negligible computational effort. 

In this paper, we implement this simple CR scheme into the \texttt{AREPO} moving-mesh code \citep{springel2010}, coupling it to the IllustrisTNG galaxy formation model \citep{weinberger2017,pillepich2018}. We then run a suite of (25\,Mpc\,h$^{-1}$)$^3$ cosmological magnetohydrodynamical boxes, at TNG100 resolution, and study the impact of CRs on galaxy and halo properties. The paper is organised as follows: in Section~\ref{sec:methods}, we provide an overview of the simulation suite and methods used. Results are presented in Section~\ref{sec:results} and summarised in Section~\ref{sec:conc}.


\section{Methods}\label{sec:methods}

\subsection{Initial Conditions and Setup Overview}

Results presented in this work are derived from a suite of (25\,Mpc\,h$^{-1}$)$^3$\,$\sim$\,(37\,Mpc)$^{3}$ cosmological magnetohydrodynamical boxes run with the moving mesh code \texttt{AREPO} \citep{springel2010}. The initial conditions, generated using \texttt{N-GenIC} \citep{springel2005}, are identical to the large number of TNG variation boxes presented in \cite{pillepich2018}. In particular, we use the L25n512 box containing 512$^3$ dark matter particles and 512$^3$ initial gas cells, giving a dark matter (average baryonic) mass of $\sim$\,10$^7$ (10$^6$)\,M$_\odot$. This corresponds to a spatial resolution of $\sim 350$\,pc (the average star-forming gas cell size) to $\sim 740$\,pc (the gravitational softening of stars and dark matter), equivalent to the TNG100-1 simulation \citep{nelson2019b}.

Due to the relatively small size of the simulation domain, massive objects (M$_{\rm{200c}}$\,$\gg$\,$10^{12}$\,M$_\odot$) are limited in number. At the other end of the mass spectrum (M$_{\rm{200c}}$\,$\ll$\,$10^{10.5}$\,M$_\odot$), structures are typically sampled by $\lesssim$\,$O(1000)$ baryonic elements. In order to maximise statistics and resolution, we focus our main analysis on halos in the mass range $\sim$\,$10^{11}$\,-\,$10^{12}$\,M$_\odot$, identified via the friends-of-friends algorithm \citep{davies1985}. In addition, unless otherwise stated, we restrict the analysis to central galaxies of these haloes, as identified by \texttt{SUBFIND} \citep{springel2001}. Throughout this work, the term halo mass refers to M$_{\rm{200c}}$, and (galaxy) stellar mass to the cumulative mass of all stars within twice the stellar half mass radius.

We adopt a cosmology consistent with the Planck 2016 analysis \citep{planck2016}: $\Omega_\Lambda = 0.6911$, $\Omega_{\rm m} = 0.3089$, $\Omega_{\rm b} = 0.0486$ and $h = 0.6764$ [100 km s$^{-1}$ Mpc$^{-1}$].

\subsection{Galaxy Formation Model}\label{ssec:methods_model}

We use the IllustrisTNG model (hereafter, TNG model) for the physics of galaxy formation and evolution \citep{weinberger2017,pillepich2018}. TNG includes models for primordial and metal line cooling in the presence of a time-dependant metagalactic radiation field, formation of stars and supermassive black holes (SMBH), stellar evolution and enrichment, feedback processes driven by both supernovae (SNe) and SMBHs, and other related processes. The TNG model also includes ideal magnetohydrodynamics \citep{pakmor2011,pakmor2014} with an initial ($z$\,$\sim$\,$127$) primordial seed field of $\sim$\,$10^{-14}$\,cG, while the \citealt{powell1999} eight-wave divergence cleaning scheme is used to maintain $\nabla \dcdot \vec{B}$\,$\sim$\,$0$.

In this work, we freeze every aspect of the TNG model but one: stellar winds, i.e. the deposition of mass, energy and momentum transfer from SNe feedback \citep{springel2003}. These are given 90$\%$ of their original (kinetic) energy, while the remaining 10$\%$ is assumed to accelerate CRs \citep[e.g.][]{helder2012,ackerman2013}. At all other positions in space, we estimate the energy density of CRs ($e_{\rm{cr}}$) using the \cite{hopkins2023b} sub-grid model. This adopts a number of approximations, including but not limited to: assuming spherical symmetry around point-like sources, constant streaming/diffusion coefficients in space and time, and that the CR energy equation is in a steady-state. Ultimately,  $e_{\rm{cr}}$ at the location of a gas cell in the simulation $i$ is estimated as the sum over all ($j$) star forming gas cells: $e_{\rm{cr,i}} = Q^{\rm{atn}}_{i}\, \sum_{j}\, \dot{E}_{\rm{cr,j}}^{\rm atn}\,F(r_{ij})$, where $Q^{\rm{atn}}_{i} = e^{-\Delta \tau_{\rm{cr,i}}}$ accounts for the attenuation of CRs. Here, 
\begin{equation}
\Delta\tau_{\rm{cr,i}} = \frac{\psi_{\rm{loss}}^{i}}{2}\, \left[ \Delta x_{i}^{2} + \left( \frac{\rho_{{\rm{gas,i}}}}{|\nabla \rho_{{\rm{gas,i}}}|}  \right)^{2} \right]^{1/2}
\end{equation}
incorporates local gas properties. $\psi_{\rm{loss}}$\,$\sim$\,$10^{-16}$\,(6.4\,$n_{\rm n}$ + 3.1\,$n_{\rm e}$ + 1.8\,$n_{\rm HI}$)\footnote{$n_{\rm n}$, $n_{\rm e}$ and $n_{\rm HI}$ are the number densities of nucleons, electrons and neutral atoms.}\,cm$^3$\,$s^{-1}$ represents a cumulative loss term for various dissipation processes \citep{mannheim1994,guo2008}, while $\Delta x$, $\rho$ and $\nabla \rho$ correspond to the size, density and density gradient of the corresponding gas cell.

The energy input rate is $\dot{E}_{{\rm{cr,j}}}^{\rm{atn}}= Q^{\rm{atn}}_{j} \,\langle \dot{E}_{\rm{cr}} \rangle_{j}$, where the (attenuated) contribution by a star-forming gas cell $j$ is given by $\langle \dot{E}_{\rm cr} \rangle_{j} = \epsilon_{\rm cr}^{\rm SNe}\,E^{\rm SNe}\,\dot{M}_{j}$. Here, $\epsilon_{\rm cr}^{\rm SNe}$ is the fraction of SNe energy transferred to CRs, which we fix at 10$\%$. An energy input of $E^{\rm SNe}$ is available for every unit stellar mass formed, which for the initial mass function and SNe progenitor mass we employ is $\sim$\,$5.2$\,$\times$\,$10^{49}$\,erg\,M$_\odot$$^{-1}$, and $\dot{M}$ is the star formation rate. In effect, this uses a `Locally Extincted Background Radiation in Optically-thin Networks' (LEBRON) type approach \citep{hopkins2018}, where the attenuation over the entire trajectory from the source to the target is approximated by the local terms at these locations.

Lastly, $F(r_{ij}) = [4\pi\,r_{ij}\,(\kappa_{\rm{eff}} + v_{\rm st,eff}\,r_{ij})]^{-1}\,e^{-r_{ij}^{2}/r_{\rm max}^{2}}$ modulates the $e_{\rm{cr}}$ contribution based on the distance ($r_{ij}$) between the source and the target. $\kappa_{\rm{eff}}$ and $v_{\rm st,eff}$ are the two free parameters of the model that correspond to effective diffusion- and streaming-like terms. The scale $r_{\rm max}$\,$=$\,$(t_{\rm max}\,v_{\rm st,\,eff} / 2)\,\{1 + [1 + 16\,\kappa_{\rm eff} / (v_{\rm st,\,eff}^{2}\,t_{\rm max)} ]^{1/2} \}$ is a proxy for the maximum distance that CRs may possibly propagate within a time interval $t_{\rm max}$, which at each timestep we set to the time since the start of the simulation (see \citealt{hopkins2023b} for a more detailed discussion).

At every timestep, $e_{\rm{cr}}$ is computed for all active gas cells using a tree-walk similar to the one often used for gravity \citep{barnes1986}. Following the calculation, the corresponding (non-thermal) pressure contribution [P$_{\rm cr}$ $=$\,$(\gamma_{\rm cr} - 1) e_{\rm{cr}}$\,$=$\,$e_{\rm{cr}}/3$] is added under the ultra-relativistic approximation ($\gamma_{\rm cr}$\,$=$\,$4/3$), as well as a heating term [$\dot{e}_{\rm th,\,gas} = e_{\rm cr}\,(0.9\,n_{n} + 1.6\,n_{e})\times10^{-16}\,{\rm cm^{3}}\,{\rm s^{-1}}$] that accounts for Coulombic and hadronic losses \citep{mannheim1994,guo2008}, and an ionization rate [$\Gamma_{\rm cr} = 7.5\times 10^{-18}\,{\rm s^{-1}}\,(e_{\rm cr}/{\rm eV\,cm^{-3}})$] due to low-energy non-relativistic CRs \citep{indriolo2015}. 

Throughout this work, we only consider SNe (in practice, star-forming gas cells) as CR sources. While one could additionally add analogous source terms arising from e.g. relativistic jets driven by AGN, or acceleration in cosmic shocks, we forego these contributions in this preliminary study. Such contributions will be important in massive elliptical galaxies where black holes drive powerful jets \citep[e.g.][]{omma2004,teyssier2011}, and can particularly dominate the CR injection fraction in the absence of young populations of massive stars. For this reason, we restrict our current analysis to Milky Way-mass halos and below, with M$_{\rm{200c}}$\,$\lesssim$\,$10^{12}$\,M$_\odot$.

In the main body of work we also intentionally neglect CR cooling terms due to the excitation of Alfv\'en-waves \citep{wiener2017,ruszkowski2017,buck2020}, i.e. Alfv\'en heating of the gas. Appendix~\ref{app:alfven_heating} demonstrates that this process is negligible in our simulations.


\section{Results}\label{sec:results}

\subsection{Selecting a Fiducial Set of Parameters}\label{sec:param_var}

\begin{figure*}[ht!]
    \centering
    \includegraphics[width=18cm]{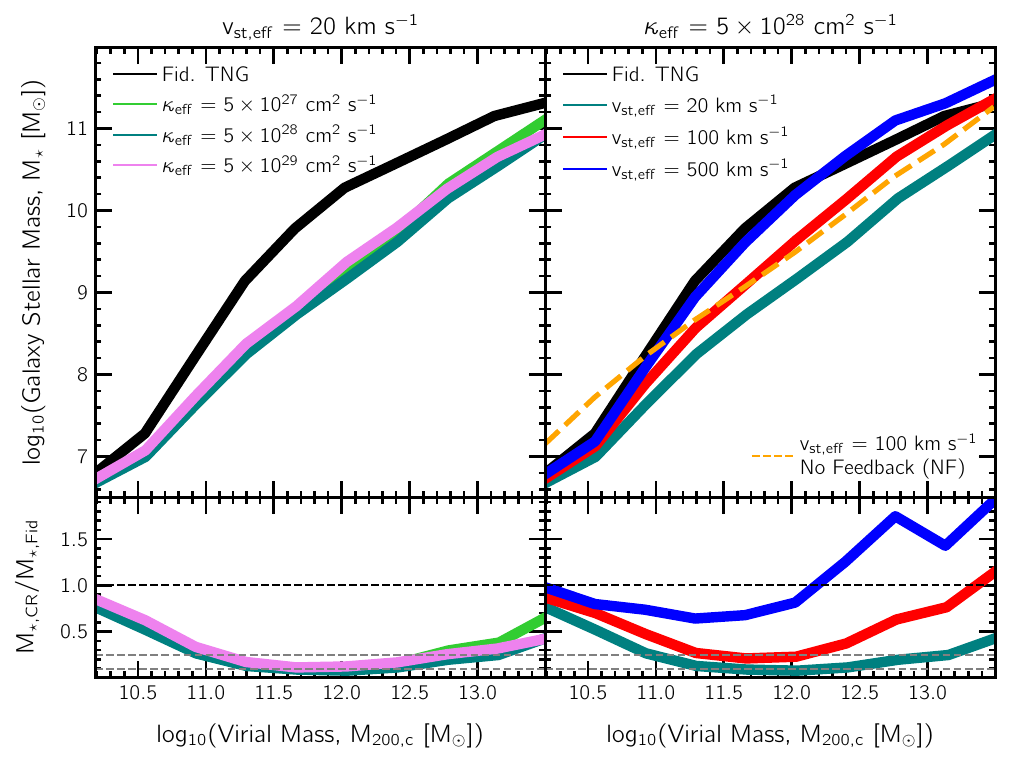}
    \caption{The impact of changing $v_{\rm{st,eff}}$ and $\kappa_{\rm{eff}}$, the two free parameters of the CR model, on the the stellar-to-halo mass relation at $z$\,$=$\,$0$. We vary $\kappa_{\rm{eff}}$ at fixed $v_{\rm{st,eff}}$ (left panels), and vice versa (right panels). The black curve shows the relation from the realisation with the fiducial TNG model (no CRs), while the various colored curves are simulations with CRs included. The bottom panels show the ratio of the various CR cases to the TNG model, where the dashed black (gray) horizontal lines show values of one (0.25 and 0.1). Most models including CRs have significantly suppressed stellar mass growth. Varying $\kappa_{\rm{eff}}$ at fixed $v_{\rm{st,eff}}$ has a minimal impact, while larger values of $v_{\rm{st,eff}}$ at fixed $\kappa_{\rm{eff}}$ lead to higher stellar masses, closer to TNG values. Throughout the paper, we adopt values of $100$\,km\,s$^{-1}$ ($5 \times 10^{28}$\,cm$^2$\,s$^{-1}$) $v_{\rm{st,eff}}$ ($\kappa_{\rm{eff}}$) as our fiducial set of transport parameters (red curve, labeling it `TNG-CR'). For this case, we also contrast against a simulation with no SNe/SMBH feedback (NF; dashed orange curve). Finally, for comparison in subsequent plots, we also consider a weaker model (the blue curve, labeling it `TNG-Weak CR'). }
    \label{fig:param_space_var}
\end{figure*}

We begin with a brief exploration of $v_{\rm{st,eff}}$ and $\kappa_{\rm{eff}}$ -- the effective streaming velocity and diffusion coefficient -- that are the two free parameters of the cosmic ray transport model. Fig.~\ref{fig:param_space_var} explores this parameter space in terms of the resulting stellar-to-halo mass relation at $z$\,$=$\,$0$. Black curves show the median relation from the realisation with the fiducial TNG model, which is in reasonable agreement with observational/empirical constraints \citep[e.g.][]{behroozi2013,moster2013}. Analogous median relations from the various CR runs are shown through the various colored curves. The bottom panels show the ratio of the various CR cases to the TNG model, i.e. the deviation of the (median) stellar mass, at fixed halo mass. 

To choose values for these two CR parameters, we first draw inspiration from the calibration of \cite{hopkins2023b}. By contrasting a set of properties against simulations run using an explicit CR transport scheme with a constant anisotropic diffusivity $\kappa_\parallel$\,$=$\,$3\,\times\,10^{29}$\,cm$^2$\,s$^{-1}$ \citep{hopkins2020}, they found that values of ($v_{\rm{st,eff}}$, $\kappa_{\rm{eff}}$)\,$=$\,($20$\,km\,s$^{-1}\,,\,5\,\times\,10^{28}$\,cm$^2$\,s$^{-1}$) offer a good fit. We therefore explore variations about this starting point, since they are in agreement with observationally favoured values (see discussion below).

On the left panels, we fix $v_{\rm{st,eff}}$\,$=$\,$20$\,km\,s$^{-1}$ and vary $\kappa_{\rm{eff}}$ by an order of magnitude in each direction. The median curves corresponding to $\kappa_{\rm{eff}}$\,$=$\,$5\,\times\,[10^{27}, 10^{28}, 10^{29}]$\,cm$^2$\,s$^{-1}$ are shown in green, teal and violet, respectively. In all cases, a clear suppression in stellar mass is seen at fixed halo mass. Starting from a value of $\sim$\,$0.8$ at M$_{\rm{200c}}$\,$\sim$\,$10^{10}$\,M$_\odot$, the ratio (lower panel) reaches a minima of $\sim$\,$0.1$ at the Milky Way-mass (M$_{\rm{200c}}$\,$\sim$\,$10^{12}$\,M$_\odot$). A similar qualitative behaviour was seen in \cite{hopkins2020}, i.e. in simulations run with a more advanced CR model, where stellar masses were suppressed only in halos above a mass M$_{\rm{200c}}$\,$\gtrsim$\,$10^{11}$\,M$_\odot$.

Unexpectedly, the three curves are similar over almost the entire range of M$_{\rm{200c}}$, i.e. varying $\kappa_{\rm{eff}}$ by two orders of magnitude has minimal impact on the stellar-to-halo mass relation, except for the high-mass end (M$_{\rm{200c}}$\,$\gtrsim$\,$10^{13}$\,M$_\odot$) where the red curve begins to separate out from the others. While \cite{hopkins2020} also noted that the diffusion coefficient has a minimal impact on stellar mass growth at masses M$_{\rm{200c}}$\,$\lesssim$\,$5\,\times\,10^{11}$\,M$_\odot$, they found that larger values of $\kappa_\parallel$ led to greater suppression in stellar mass for more massive halos (M$_{\rm{200c}}$\,$\gtrsim$\,$5\,\times\,10^{11}$\,M$_\odot$). The difference must arise from the simplicity of the CR model we use and/or differences in other baryonic physics models or numerical schemes between the simulations.

In the right panel of Fig.~\ref{fig:param_space_var}, we fix $\kappa_{\rm{eff}}$\,$=$\,$5\,\times\,10^{28}$\,cm$^2$\,s$^{-1}$, and vary $v_{\rm{st,eff}}$\,$=$\,$[20,\,100,\,500]$\,km\,$s^{-1}$, shown in teal, red and blue, respectively. Although stellar masses (at fixed halo mass) are once again displaced with respect to the black curve, a monotonic trend is visible: the level of suppression is reduced towards larger values of $v_{\rm{st,eff}}$, particularly at the low mass end. The red curve ($v_{\rm{st,eff}}$\,$=$\,$100$\,km\,$s^{-1}$) returns stellar masses within a factor of $\lesssim$\,$4$ of the fiducial TNG result at all masses, with the difference maximal at the Milky Way-mass range. An even larger $v_{\rm{st,eff}}$ of $500$\,km\,$s^{-1}$ brings the median to within a factor of $\lesssim$\,$1.5$ at the low mass end (M$_{\rm{200c}}$\,$\lesssim$\,$10^{12}$\,M$_\odot$), and in fact over-estimates the stellar mass in more massive halos. As we discuss below, this is due to delayed SMBH growth, making the AGN feedback model of TNG relatively less effective above the Milky Way-mass range.

It is encouraging that larger values of $v_{\rm{st,eff}}$ produce stellar-to-halo mass relations similar to TNG, as such a model can likewise be consistent with observational constraints. However, it is unclear if these CR transport parameters are physically plausible. \cite{hopkins2021} suggest that the narrow range in parameter space defined by 
$\kappa_{\rm{eff}} + v_{\rm{st,eff}}\,r_{\odot}^{\rm gal} \sim O(10^{29}\,{\rm cm^{2}\,s^{-1}})$, where $r_{\odot}^{\rm gal} \sim 8\,$kpc is the galactocentric distance of the sun, is needed to be consistent with observations in the solar neighbourhood \citep[see also][]{korsmeier2022}.

Irrespective of the value assumed for $\kappa_{\rm{eff}}$, an effective streaming speed $v_{\rm{st,eff}}$\,$=$\,$500$\,km\,s$^{-1}$ is on the higher side ($v_{\rm{st,eff}}$\,$r_{\odot}^{\rm gal}$\,$\sim$\,$10^{30}$\,cm$^{2}$\,s$^{-1}$). A value of $v_{\rm{st,eff}}$\,$=$\,$100$\,km\,s$^{-1}$, paired with either $\kappa_{\rm{eff}}$\,$=$\,$5\,\times\,10^{27}$ or $5\,\times\,10^{28}$\,cm$^2$\,s$^{-1}$, satisfies the above constraint. Both values of $\kappa_{\rm{eff}}$ yield largely similar results, and we adopt the latter throughout the rest of this work for computational convenience, i.e. we fix $\kappa_{\rm{eff}}$\,$=$\,$5\,\times\,10^{28}$\,cm$^2$\,s$^{-1}$ and $v_{\rm{st,eff}}$\,$=$\,$100$\,km\,s$^{-1}$, and label this `TNG-CR'. 

In order to offer a comparison with a case where the stellar-to-halo mass relation is more realistic, we also include a subset of results in a number of plots from the run with $v_{\rm{st,eff}}$\,$=$\,$500$\,km\,s$^{-1}$, which we label as `TNG-Weak CR'. However, we again stress that such a case is in tension with observational CR constraints  \citep{hopkins2021}, and is only meant as a proxy for a case where the impact of CRs is relatively weak.

The final comparison we make throughout this paper is a `no feedback' case. Here, we adopt the fiducial CR values of $\kappa_{\rm{eff}}$\,$=$\,$5\,\times\,10^{28}$\,cm$^2$\,s$^{-1}$ and $v_{\rm{st,eff}}$\,$=$\,$100$\,km\,s$^{-1}$, but completely disable all stellar and SMBH feedback processes. We label this `TNG-CR-NF' (dashed orange line, upper-right panel of Fig.~\ref{fig:param_space_var}). As seen, this differs from TNG-CR (red curve) only for relatively small halos (M$_{\rm{200c}}$\,$\lesssim$\,$10^{11}$\,M$_\odot$), beyond which the two are largely similar. This suggests that the suppression of stellar masses (at fixed halo mass) is dominated by the effect of CRs, and furthermore that \textit{no} recalibration i.e. weakening of the TNG feedback model could bring the results into agreement with data. We discuss the implications of this further below.

\begin{figure*}[ht!]
    \centering
    \includegraphics[width=18cm]{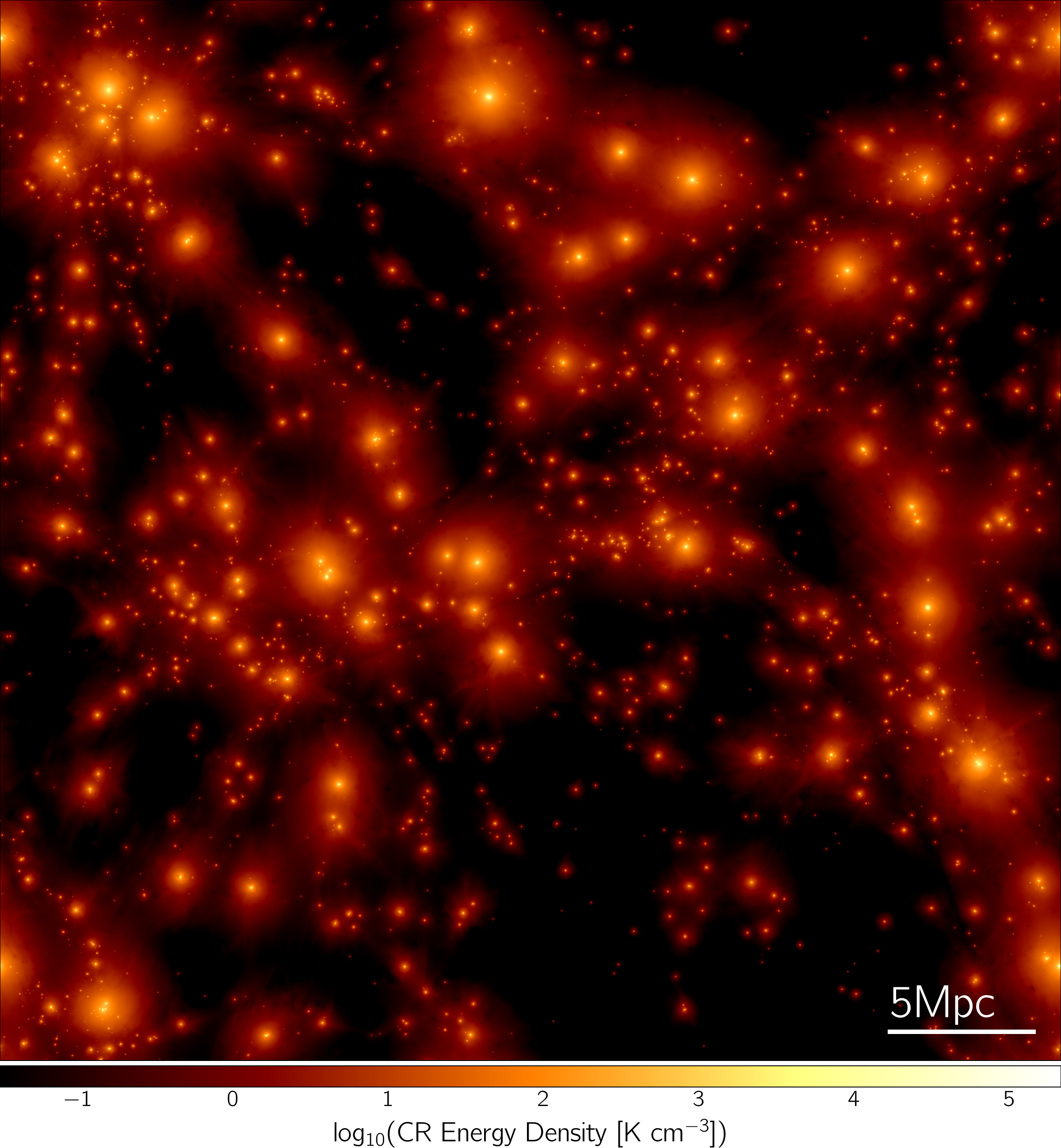}
    \caption{A visualisation of the projected CR energy density through the simulated volume (at $z$\,$=$\,$0$) of the realisation with ($v_{\rm{st,eff}}$, $\kappa_{\rm{eff}}$)\,$=$\,($100$\,km\,s$^{-1}\,,\,5\,\times\,10^{28}$\,cm$^2$\,s$^{-1}$), i.e. the fiducial set of parameters we use. The image extends 25\,h$^{-1}$\,$\sim$\,37\,Mpc along the x- and y-axes of the plane, as well as in the projection direction. CR energy densities are as high as $\gtrsim$\,$10^5$\,K\,cm$^{-3}$ at the centres of galaxies, i.e. at the sites of star formation, and decay with increasing distance. Halos with multiple such sources, i.e. contributions from satellite galaxies in addition to the central, have multiple bright `dots', leading to a certain degree of asymmetry in the CR energy density profile; the halo towards the lower-left of the image is a representative example. The corresponding non-thermal CR pressure, as well as secondary effects due to associated heating+ionization terms, has a noticeable impact on the evolution of galaxies, as we shall explore throughout the rest of this work.}
    \label{fig:mainVisImage}
\end{figure*}

Before expanding our view on the statistics of galaxy properties across the population, we consider the spatial distribution of cosmic rays on cosmological scales. Fig.~\ref{fig:mainVisImage} shows a visualisation\footnote{In all visualisations presented in this work, colors show the mass weighted mean of the corresponding quantity along the projection direction.} of the large-scale structure of the CR energy density ($e_{\rm{cr}}$) across the simulated volume of the TNG-CR realisation. Brighter colors correspond to larger $e_{\rm{cr}}$, as shown by the colorbar. CR energy densities are as high as $\gtrsim$\,$10^5$\,K\,cm$^{-3}$ at the centers of galaxies, corresponding to sites of active star formation. CR energy decays as distance increases, in a roughly spherical symmetric fashion owing to the functional form of the distance-modulation factor ($F(r_{ij})$; Section~\ref{ssec:methods_model}). This gives rise to cosmic ray halos that permeate throughout, and extend beyond, dark matter halos.

Halos with multiple sites of star formation, i.e. those that have star-forming satellites in addition to a star-forming central, have multiple such halo features. The superposition of these gives rise to a certain degree of asymmetry in the $e_{\rm{cr}}$ structure. The halo towards the lower-left of the image is a representative example. In addition, complexity in the gas distribution in and around galaxies, i.e. in $\rho$ and $\nabla \rho$, also lead to deviations from perfectly spherical $e_{\rm{cr}}$ profiles ($\Delta \tau_{\rm cr,i}$; Section~\ref{ssec:methods_model}). Such features are most clearly visible towards the outer edge of halos, sometimes as linear or filamentary-like streaks between halos. The corresponding non-thermal pressure, as well as secondary effects due to associated heating and ionization terms, has a noticeable impact on the evolution of galaxies, as we shall explore throughout the rest of this work.

\subsection{Integrated and Large-Scale Halo Properties}

\begin{figure*}
    \centering
    \includegraphics[width=16.5cm]{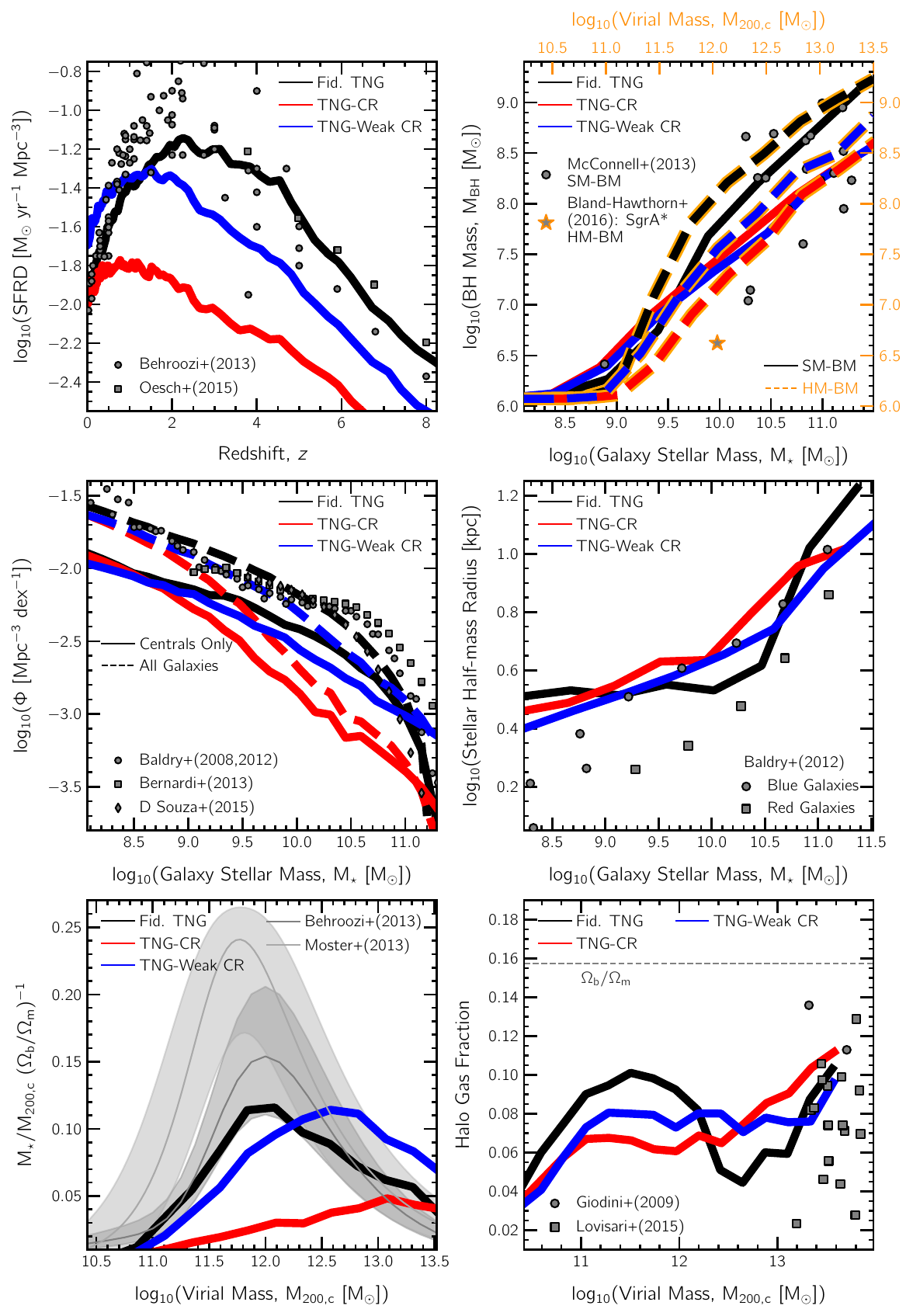}
    \caption{A summary of galaxy and halo scale integrated properties of the galaxy population, all at $z$\,$=$\,$0$ but for the top-left. \textit{Caption continued below.}}
    \label{fig:prop_comp}
\end{figure*}

\addtocounter{figure}{-1}
\begin{figure*}[ht!]
    \centering
    \caption{\textit{continued} In all panels, the black curves correspond to the fiducial TNG model, while the red (blue) curves to TNG-CR (TNG-Weak CR). Scatter points show various observational and empirical constraints: \protect{\citealt{baldry2008,giodini2009,baldry2012,bernardi2013,behroozi2013,mcconnell2013,moster2013,dsouza2015,oescg2015,lovisari2015,blandhawthorn2016}}. The fiducial TNG model is in reasonable agreement with these observational constraints, as this is the set of observables used for calibration. The TNG-CR simulation however produces a reduced number of galaxies at $M_\star$\,$\gtrsim$\,$10^{8.5}$\,M$_\odot$ (center left) as a result of lower star formation rate densities over cosmic epochs (top left). Despite the suppressed stellar mass at fixed halo mass (lower left), and lower black hole masses in general (top right), the galaxy size-stellar mass relation (center right) and halo gas fractions (lower right) are within a factor of $\sim$\,few between TNG-CR and the fiducial TNG case, as well as the different data points (see main text for a more detailed discussion).}
\end{figure*}

Fig.~\ref{fig:prop_comp} shows six properties and relations of the galaxy population: the star formation rate density, stellar mass function, stellar mass to halo mass relation, black hole mass versus stellar mass, stellar sizes, and halo gas fractions. Black curves always show the fiducial TNG model, contrasting against TNG-CR (red) and TNG-Weak CR (blue). Gray points and shaded bands show various observational and empirical constraints. Note that these are exactly the observables used in the calibration of the TNG model, undertaken in the relative sense with respect to the outcome of the original Illustris simulation \citep{pillepich2018}. We therefore reflect on how TNG, in its `calibration space', is impacted by the inclusion of cosmic rays. Except for the top left panel, all curves show the median, and all give $z=0$ results.

The top-left panel shows the cosmic star formation rate density (SFRD) as a function of redshift. At $z$\,$\gtrsim$\,$2$, the SFRD in the TNG-CR case (red) is suppressed by a factor $\sim$\,$4-6$ with respect to TNG (black), with the disparity between the two growing smaller at $z$\,$\lesssim$\,$2$. The inclusion of CRs shifts the peak of the SFRD to lower redshift, while the shape at high redshift is similar. Coincidentally, the $z$\,$=$\,$0$ SFRD of TNG-CR reaches the same value of TNG. Neglecting all possible complications including obscuration effects, both CR models shown are in significant tension with data \citep{behroozi2013,oescg2015}.

The reduced global star formation rates naturally lead to galaxies being typically under-massive in their stellar content, which we show through the galaxy stellar mass function ($\Phi$, middle left panel). Solid curves show $\Phi$ for central galaxies, i.e. the galaxies we primarily focus on in this work, and dashed curves for all galaxies, i.e. centrals and satellites. In both these sets of populations, the number density of M$_\star$\,$\gtrsim$\,$10^{8.0-8.5}$\,M$_\odot$ galaxies is lower in TNG-CR (red) versus TNG (black). The TNG-Weak CR case typically lies in between these two, except for $z$\,$\lesssim$\,$1$ where the SFRD is larger. An imprint of this enhanced star formation activity is visible as a larger number density of M$_\star$\,$\gtrsim$\,$10^{11.0}$\,M$_\odot$ galaxies in the centre-left panel. Broadly speaking, the dashed all galaxies lines should be compared against the observational constraints \citep{baldry2008,bernardi2013,dsouza2015}. While all models are roughly consistent with data in the dwarf regime, both CR models are ruled out by the observed abundance of galaxies at the knee i.e. Milky Way-mass systems.

For a given halo mass, the inclusion of CRs has a noticeable impact on the stellar mass content of galaxies. The lower left panel shows the stellar-to-halo mass relation at $z=0$, now normalizing stellar mass by halo mass. The TNG-CR simulation clearly has a different behavior when compared to TNG (black), both in terms of shape and normalisation. A peak, if it exists, is marginal and moved to roughly one order of magnitude higher in halo mass. In contrast, TNG-Weak CR (blue) is qualitatively similar to TNG, but horizontally offset by $\sim$\,$0.5$\,dex. Neither are consistent with semi-empirical models \citep{behroozi2013,moster2013}.

In addition to the reduction of stellar mass, the growth of supermassive black holes is also hindered. The black hole versus halo mass relation is shown in the top right panel with the dashed curves outlined in orange, with x-axis values (i.e. halo mass) given by the top (orange) x-axis. At M$_{\rm 200c}$\,$\gtrsim$\,$10^{11}$\,M$_\odot$, TNG-CR is offset by $\sim$\,$0.5$\,dex below TNG, signifying smaller black holes at fixed halo mass. Interestingly, the median relation in TNG-CR falls closer to the mass of SgrA* in the Milky Way \citep{blandhawthorn2016}, shown through the gray star with an orange outline, which is otherwise difficult to reproduce in MW-mass galaxies \citep{pillepich2023}. Once again, the TNG-Weak CR case lies in between the fiducial TNG and TNG-CR outcomes, as expected.

The upper-right panel also shows the black hole-to-stellar mass relation, with the solid curves and lower (black) x-axis. In this case, the TNG and TNG-CR models vary only by a factor of a $\sim$\,few, and are both in broad agreement with the observed relation reported by \citealt{mcconnell2013}. The TNG-Weak CR case is largely similar to TNG-CR in this case, i.e. at fixed stellar mass, the mass of the black hole is largely independent of $v_{\rm{st,eff}}$. This is expected from the functional form of $F(r_{ij})$ (Section~\ref{sec:methods}), since the distance modulation factor is largely independent of $v_{\rm{st,eff}}$ at small $r_{ij}$, and thus has a minimal impact on the accretion rate of black holes, which is rather modulated by the density of the local gas reservoir.

The centre right panel shows the galaxy size-mass relation, with the former defined as the stellar half mass radius. All three models are largely similar, and all broadly consistent with observations \citep{baldry2012}. At these masses, the population is dominated by blue galaxies, and so best compared to the circular gray data points. Although galaxies are under-massive for their parent halo mass in both TNG-CR and TNG-Weak CR, the size comparison demonstrates that some galaxy properties are realistic i.e. largely unchanged, for a given galaxy stellar mass.

Lastly, we show the halo gas fraction as a function of the halo mass in the lower right panel. To guide the eye, the dashed gray line is drawn at the cosmic baryon fraction, i.e. $\Omega_{\rm{b}}$/$\Omega_{\rm{m}}$\,$\sim$\,$16\%$. While the three curves vary by a few percent at halo masses M$_{\rm{200c}}$\,$\lesssim$\,$10^{13}$\,M$_\odot$, they are in good agreement at higher masses. For simplicity we plot the the FoF gas mass normalized by the total FoF mass, which can only roughly be compared to observational constraints \citep{giodini2009,lovisari2015}, against which there is reasonably agreement in all models. Analogous observations at lower mass scales will play an important role in discriminating between various galaxy formation models in the future \citep{zhang2024}.

\begin{figure*}[ht!]
    \centering
    \includegraphics[width=18cm]{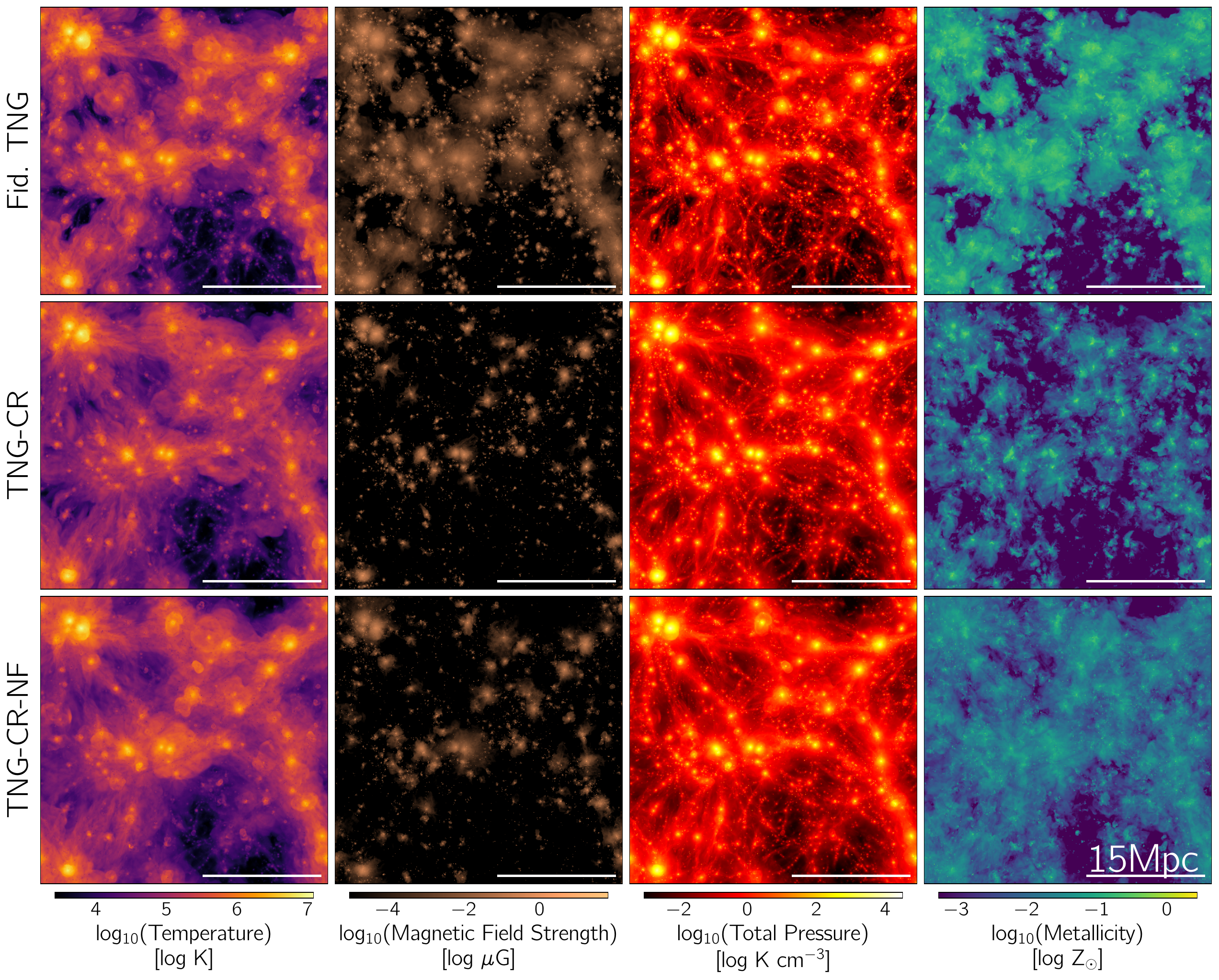}
    \caption{A visual impression of a subset of gas quantities through the simulated volumes for the fiducial TNG (upper panels), TNG-CR (centre panels) and TNG-CR-NF (lower panels) runs. From left to right, we show temperature, magnetic field strength, (total) pressure, and metallicity. All images extend $\sim$\,$37$\,Mpc along the plane and the perpendicular direction, i.e. through the entire ($z$\,$=$\,$0$) simulated volume, and scale bars in all panels correspond to $15$\,Mpc. The large-scale structure of temperature and pressure is similar between the three runs, but for the lack of hot, rarified `bubbles' around massive galaxies in TNG-CR. The runs differ significantly in the other quantities, with lower levels of magnetisation and metal enrichment of halo gas in the two CR runs as compared to TNG. The similarity in total pressure despite weaker field strengths, implying lower levels of magnetic pressure, shows that the non-thermal support by CRs is significant.}
    \label{fig:fid_vs_cr_image}
\end{figure*}

Having focused on integral galaxy population statistics and scaling relations, we now transition to focus on the properties of gas in and around galaxies, to better understand the physical impact of CRs. We begin with a visual motivation in Fig.~\ref{fig:fid_vs_cr_image}, which contrasts projections of four quantities in the fiducial TNG case (upper panels) versus TNG-CR (center panels) and TNG-CR-NF (no feedback; lower panels). From left to right, we show temperature, magnetic field strength, (total) pressure, i.e. thermal plus non-thermal, and metallicity.

On these large scales, the distributions of temperature and total pressure are largely similar between the three runs, except for the lack of hot rarified `bubbles' around massive halos (M$_{\rm{200c}}$\,$\gtrsim$\,$10^{12}$\,M$_\odot$) in TNG-CR. In the (fiducial) TNG runs, these are understood to be produced by kinetic mode AGN feedback \citep{pillepich2021}, the dominant channel for massive black holes (M$_{\rm{BH}}$\,$\gtrsim$\,$10^8$\,M$_\odot$; \citealt{weinberger2017}). Given that black holes are less massive at fixed halo mass in the CR runs, these features are therefore suppressed in the lower panels. 

We remark on an interesting qualitative result. Although stellar and black hole feedback processes are turned off in TNG-CR-NF, and thus it contains no AGN driven outflows, bubbles similar to TNG are visible in this case. We speculate that this reflects outflows driven purely by CR pressure gradients, a phenomenon previously seen in idealized simulations and on order of magnitude smaller spatial scales \citep{hanasz2013, salem2014,girichidis2016,huang2022}, and in the FIRE-2 cosmological simulations on Mpc scales \citep{hopkins2021b}.

The similarity of the (total) pressure images is surprising, since the large-scale magnetic field strengths are notably different between the two runs. While the centres of halos have similar levels of magnetisation, halo gas surrounding these galaxies has much weaker field strengths. Despite the significant drop in magnetic pressure (P$_{\rm{B}}$\,$\sim$\,$|\vec{B}|^2$), the total pressure is nearly unchanged. This occurs because cosmic rays instead provide a similar level of non-thermal support, i.e. cosmic ray pressure naturally compensates for the missing magnetic pressure. 

Gas in the circumgalactic medium of galaxies also has much lower metallicity in the CR model runs. This is a natural consequence of the lower stellar masses of galaxies, and thus overall less efficient production of metals across cosmic time. As we discuss below, lower feedback-driven outflow velocities in the CR simulations also prevent metals from being distributed to as large of spatial scales into the intergalactic medium, instead being confined closer to their production sites.

\begin{figure*}[ht!]
    \centering
    \includegraphics[width=18cm]{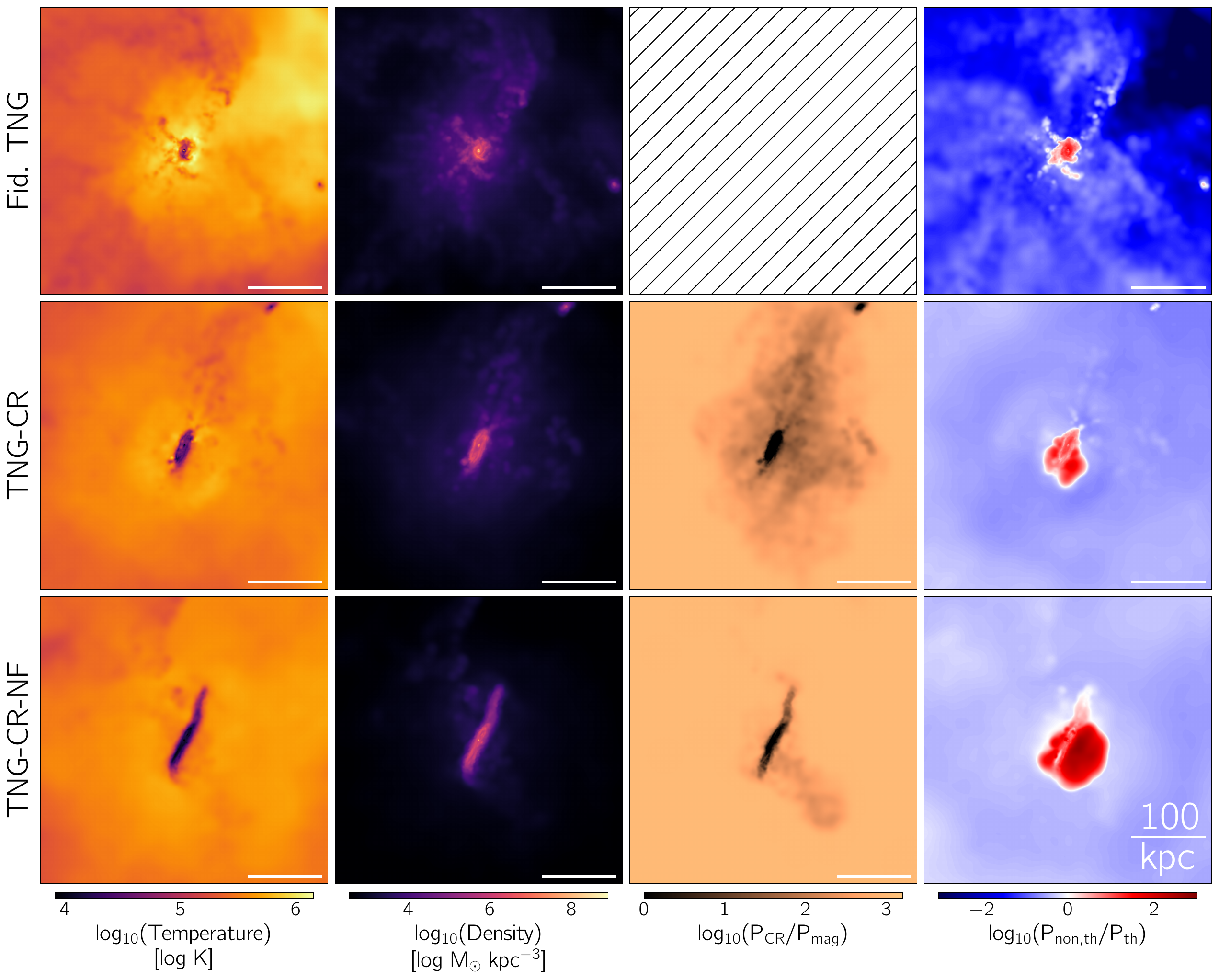}
    \caption{A visual impression of a subset of gas quantities around a Milky Way-mass halo in the fiducial TNG (upper panels), TNG-CR (centre panels) and TNG-CR-NF (lower panels) runs, cross-matched across the various realisations. Images extend $\pm$\,$R_{\rm{200c}}$ in all cases, along the plane and in the perpendicular direction, and scale bars correspond to a $100$\,kpc. From left to right, we show temperature, density, ratio of P$_{\rm{CR}}$ to P$_{\rm{mag}}$, and ratio of P$_{\rm{non,th}}$(=\,P$_{\rm{CR}}$\,+\,P$_{\rm{mag}}$) to P$_{\rm{th}}$. While profiles of temperature are largely similar across the three runs, the density of gas is lower in the TNG-CR-NF run, particularly in the halo. In the two CR runs, P$_{\rm{CR}}$ dominates over P$_{\rm{mag}}$ in all regions except the central disk, where the two are in rough equipartition. In all three runs, only the central region of the halo is dominated by non-thermal pressure support. However, the fraction of non-thermal pressure in outer regions of the halo is relatively higher in the CR runs as compared to Fid. TNG.}
    \label{fig:fid_vs_cr_vs_nf_halo_image}
\end{figure*}

In Fig.~\ref{fig:fid_vs_cr_vs_nf_halo_image}, we zoom-in to smaller scales around a single Milky Way-mass halo, cross-matched across simulations. As before, we contrast the fiducial TNG run (top row) to TNG-CR (middle row) and TNG-CR-NF (no feedback; lower row). Images span $\pm$\,$\rm{R_{200c}}$ from edge to edge, and the white scale bars always indicate $100$\,kpc. From left to right, we show temperature, density, the ratio of cosmic ray to magnetic pressure, and the ratio of non-thermal to thermal pressure.

Similar to Fig.~\ref{fig:fid_vs_cr_image}, the spatial structure of gas temperature is broadly similar across the three runs: the disk is dominated by a cool $\sim$\,$10^4$\,K component, and is embedded in a halo of hot $\gtrsim$\,$10^{5.5}$\,K gas. Hotter $\sim 10^6$\,K gas is evident in the TNG model, and somewhat suppressed in the CR runs. The extent and abundance of the cool gas disk is more prominent in the CR runs, although we attribute this primarily to a timing offset between the simulations, and not to any underlying physical cause.

On the contrary, the density structure of gas clearly differs between the runs -- most notably, gas densities are lower in the TNG-CR-NF run, particularly in the halo gas surrounding the central disk. While densities in TNG-CR are also lower in comparison to the fiducial TNG run, the level of disparity is smaller.

In the CR runs, magnetic pressure is in rough equipartition with CR pressure in the disk, as indicated by the black color. However, in all other regions of the halo, magnetic fields are clearly sub-dominant to CRs, by up to a factor of $\sim$\,$10^3$. We note that this is in qualitative agreement with results from the FIRE-2 simulations \citep{hopkins2020}. Lastly, the fourth panel of Fig.~\ref{fig:fid_vs_cr_vs_nf_halo_image} shows that only the innermost regions of the halo are dominated by non-thermal pressure support, in all three runs. As one transitions farther out into the halo, thermal pressure begins to dominate, although to a much smaller degree in the CR runs as compared to fiducial TNG.

\begin{figure*}
    \centering
    \includegraphics[width=16.5cm]{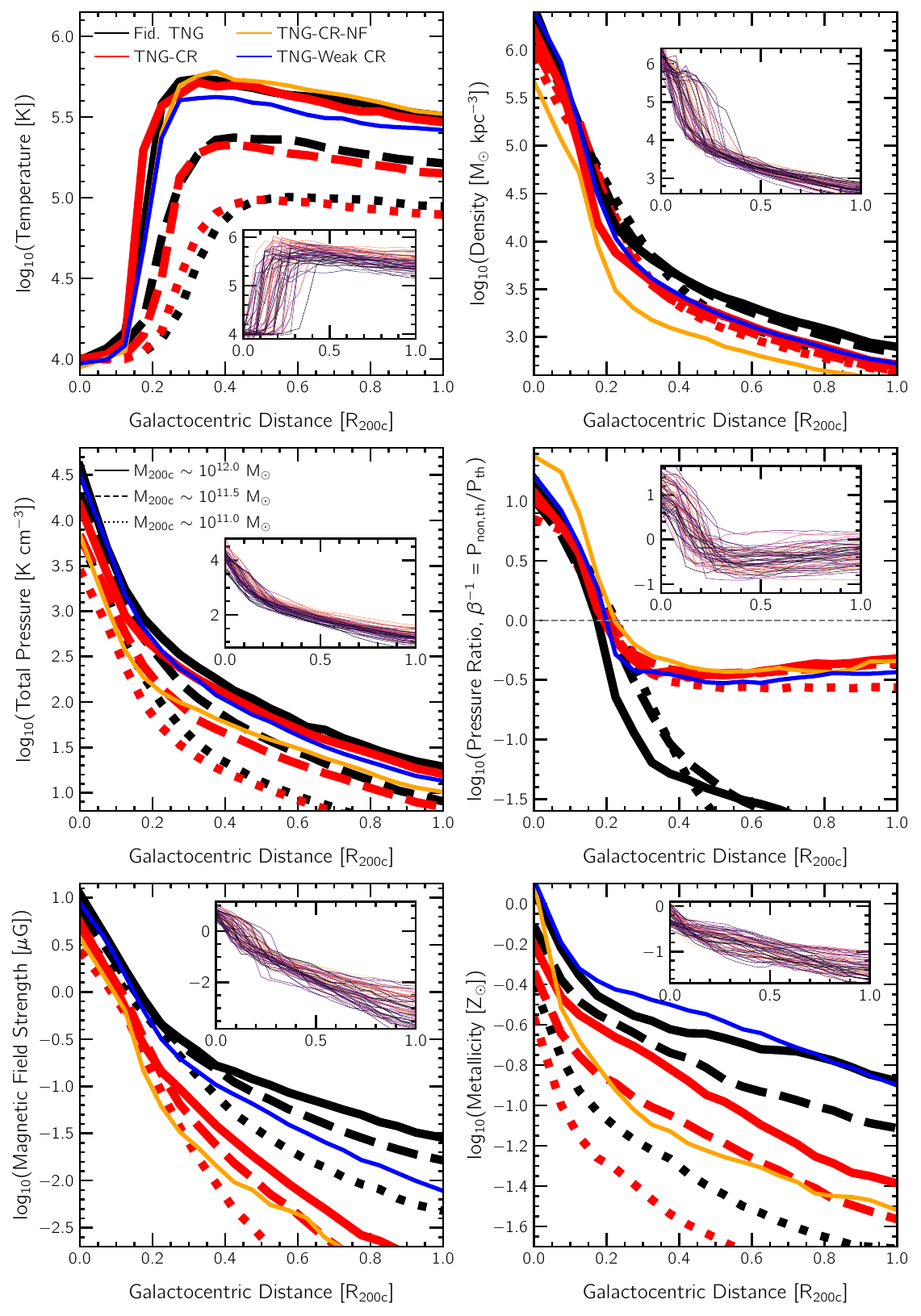}
    \caption{Spherically-averaged stacked radial profiles of six gas properties. \textit{Caption continued below.}}
    \label{fig:prof_comp}
\end{figure*}

\addtocounter{figure}{-1}
\begin{figure*}[ht!]
    \centering
    \caption{\textit{continued} Clockwise from the top left, we show: temperature, density, non-thermal to thermal pressure ratio, metallicity, magnetic field strength, and total pressure. Colors contrast TNG-CR (red) versus the fiducial TNG run (black), while the line styles correspond to different halo mass bins. For the $10^{12}$\,M$_\odot$ mass bin in each panel, we also show profiles from the CRs but no feedback run (TNG-CR-NF; orange) and TNG-Weak CR (blue). The inclusion of CRs impacts most properties: halo gas is less magnetized and metal enriched in the TNG-CR and TNG-CR-NF runs. Profiles in TNG-Weak CR are always closer to those of fiducial TNG. Most notably, the non-thermal pressure support of CGM gas ($\gtrsim$\,$0.2$\,R$_{\rm{200c}}$) is relatively larger in the CR cases as compared to the TNG run, although profiles in the galaxy ($\lesssim$\,$0.2$\,R$_{\rm{200c}}$) are largely similar across the various runs. Insets in all panels show individual profiles of Milky Way-mass halos in TNG-CR, with curves colored by halo mass (see text).}
\end{figure*}

We again note the emergence of a feature visible in TNG-CR and even in TNG-CR-NF: bubble-like outflows that are non-thermal pressure dominated. These are clearly evident in both magnetic and CR pressure, although not particularly so in temperature. The spatial scales of these bubbles is $\sim$\,$10$\,$-$\,$50$\,kpc, much smaller than the features previous discussed in Fig.~\ref{fig:fid_vs_cr_image}, and reminiscent of the Fermi/eROSITA-like bubbles seen in TNG50 \citep{ pillepich2021}. However, since they are present in the NF (no feedback) case, and thus clearly sourced by CR pressure gradients, they are of a different origin. We defer a detailed study of these bubbles and their properties to future work.

Fig.~\ref{fig:prof_comp} provides a more quantitative assessment of these trends, through spherically-averaged radial profiles of multiple gas properties. In all panels, black curves correspond to the fiducial TNG run, and red to TNG-CR. We select (central) galaxies whose parent halo masses are M$_{\rm{200c}}$\,$\sim$\,$10^{11}$, $10^{11.5}$ and $10^{12}$\,M$_\odot$, shown through dotted, dashed and solid lines, respectively. For the fiducial TNG case, note that we only select those galaxies in which the black hole in the quasar mode, thereby excluding galaxies with strong AGN feedback from our analysis. This enables us to focus on interpreting the impact of CRs with respect to star formation and stellar feedback, also facilitating comparison with previous work. All curves show median values across galaxies in the respective bin. For the Milky Way-mass bin alone (M$_{\rm{200c}}$\,$\sim$\,$10^{12}$\,M$_\odot$), we also show median curves from TNG-CR-NF (orange) and TNG-Weak CR (blue). Insets in all panels show individual profiles for Milky Way-mass halos in the TNG-CR run. Curves are colored by the virial mass of the halo, with the darkest (brightest) colors corresponding to M$_{\rm{200c}}$\,$\sim$\,$10^{11.8}$ ($10^{12.2}$) \,M$_\odot$. For lower halo masses the temperatures in the CGM are noticeably colder, which is in agreement with models of isolated halos (e.g., \citealt{girichidis2024}, see also \citealt{thomas2024}).

We begin with temperature in the top left panel. In all three mass bins, profiles are similar between TNG and TNG-CR. At the Milky Way mass, as well as the other two bins (not shown), TNG-CR-NF yields a similar radial profile, suggesting that these temperature profiles are largely modulated by gravity, and only to a secondary extent by feedback processes. Note, however, that kinetic mode feedback in TNG can produce a noticeable impact on temperature structure by driving outflows of hot, super-virial gas \citep{pillepich2021,ramesh2023a}, which we have intentionally excluded from these curves to avoid complication of interpretation of results (see above). The TNG-Weak CR curve is also similar, albeit offset to higher temperatures by $\sim$\,$0.05-0.1$\,dex.

Radial profiles of density are shown on the top-right. TNG and TNG-CR are similar in the inner halo ($\lesssim$\,$0.2$\,R$_{\rm{200c}}$), although densities are typically lower by $\sim$\,$0.1-0.2$\,dex in the latter at larger galactocentric distances. While the blue (TNG-Weak CR) solid curve lies intermediate of the black and red, the orange (TNG-CR-NF) is vertically offset further to lower values, by $0.1-0.5$\,dex depending on distance. In all the CR cases, our interpretation is that the added CR pressure inhibits gas accretion into the halo, thereby reducing total mass and overall gas densities. In the TNG-CR-NF case, star formation rates of low-mass halos (M$_{\rm{200c}}$\,$\lesssim$\,$10^{11}$\,M$_\odot$; Fig.~\ref{fig:param_space_var}) are enhanced in the absence of feedback processes, giving rise to larger CR pressures in the halo, and eventually to lower densities as they evolve into their $z$\,$=$\,$0$ descendants. We note that the trend of reduced halo gas densities in the presence of CRs is in qualitative agreement with \citealt{hopkins2020}.

In the center panels, we focus on quantities related to the pressure of gas. On the center-left, total pressure profiles are similar between the TNG and TNG-CR runs, as well as TNG-Weak CR, albeit with minor $\sim$\,$O(0.1)$\,dex variations, much like those in temperature and density. In TNG-CR-NF (orange), the total pressure is lower by $0.2-0.4$\,dex, depending on distance. This is primarily due to lower densities giving rise to lower thermal pressures, but also in part because of weaker halo magnetic fields, as we elaborate further below.

The type of pressure support, however, differs significantly between TNG and the CR runs, as shown by the ratio of non-thermal to thermal pressure (center right panel). While the profiles of TNG and TNG-CR are largely similar in the inner halo ($\lesssim$\,$0.2$\,R$_{\rm{200c}}$), where both are dominated by non-thermal pressure to similar extents, profiles begin to separate out at larger distances. In particular, the TNG ratio of $P_{\rm non,th} / P_{\rm th}$ monotonically drops as one transitions to the outer regions of the halo, to as low as $\sim$\,$0.03$ at $\sim$\,$0.6$\,R$_{\rm{200c}}$. On the contrary, the TNG-CR (and TNG-Weak CR) profiles flatten at $\sim$\,$0.3$, roughly independent of halo mass. Thus, while the total pressure support is similar between TNG and TNG-CR, thermal pressure dominates to a smaller extent in the CR runs. We mention that results from the FIRE-2 simulations are significantly different, both qualitatively and quantitatively. In FIRE-2, non-thermal pressure support dominates the CGM in halos of $\sim$\,$10^{12}$\,M$_\odot$ \citep{hopkins2020}, which is not the case in our CR runs. We make a more quantitative comparison below.

Previous studies have inferred that the CR energy density within galaxies may be in rough equipartition with analogous magnetic and thermal reservoirs \citep{cox2005,naab2017}. For our CR runs, this would imply a pressure ratio $P_{\rm non,th} / P_{\rm th}$ in the ballpark of $2$\footnote{Assuming $P_{\rm th}$\,=\,$\frac{2}{3}$\,$e_{\rm{th}}$, $P_{\rm cr}$\,=\,$\frac{1}{3}$\,$e_{\rm{cr}}$ and $P_{\rm mag}$\,=\,$e_{\rm{mag}}$.}. In all mass bins in TNG-CR, the ratio is $\gtrsim$\,$10$ at the very centres of galactic disks ($\lesssim$\,$0.05$\,R$_{\rm{200c}}$). However, in the outer disk and the inner CGM ($0.05$\,$\lesssim$\,$r$\,/\,R$_{\rm{200c}}$\,$\lesssim$\,$0.3$), the ratio is bracketed between $\sim$\,$[0.5, 10]$, i.e. an approximate equipartition is attained between thermal energy and the total non-thermal component at these intermediate distances. Further out in the halo ($\gtrsim$\,$0.1$\,$-$\,$0.2$\,R$_{\rm{200c}}$), magnetic field strengths are weak, and no longer in equipartition with other components (see discussion below). In such a case, $P_{\rm non,th}\,(\sim$\,$P_{\rm cr}) / P_{\rm th}$ is expected to be close to $0.5$ if the thermal and CR energies are in equipartition. Intriguingly, profiles of all mass bins flatten close to this value at distances $\gtrsim$\,$0.3$\,R$_{\rm{200c}}$. We suspect that this near perfect balance between the two energy reservoirs may be linked to the assumption of a steady-state $e_{\rm{cr}}$ solution by the \citealt{hopkins2023b} model (Section~\ref{ssec:methods_model}).

As discussed previously, halo magnetic fields are typically weaker in the CR runs, which we quantify further in the lower left panel of Fig.~\ref{fig:prof_comp}. Comparing the TNG and TNG-CR profiles, it is apparent that field strengths are weaker in the CR runs by as large a factor as $\gtrsim$\,$30$ in the outer reaches of the halo. The dashed and dotted lines show a similar qualitative behaviour, indicating its universality with halo mass, as to the TNG-Weak CR and TNG-CR-NF profiles, suggesting that any level of CR pressure tends to lower CGM magnetic field strengths.

In all three mass bins, B field strength is, however, largely unaffected at the very centers of galaxies. We suggest that this is because fields in the innermost regions of halos are primarily amplified by small-scale dynamos \citep{pakmor2013}. This exponential growth phase proceeds until magnetic and turbulent kinetic energy reservoirs reach rough equipartition, after which a linear growth phase takes place in the outer-disk through differential galactic rotation \citep{pakmor2017}. This explains why the TNG and TNG-CR profiles begin to differ at $\sim$\,$0.05-0.1$\,R$_{\rm{200c}}$; since galaxies are smaller at fixed halo mass in TNG-CR, this linear growth phase is effectively damped, yielding weaker fields. Finally, magnetic fields in the halo are closely linked to the transport of magnetised gas from the galaxy into the CGM \citep{pakmor2020,ramesh2023c}, suggesting that the lower level of magnetisation in the outer halo in TNG-CR is likely a result of reduced mass outflow rates.

A similar trend is seen in the case of metallicity profiles (lower right panel), where the metal enrichment of the gas in the halo outskirts is lower by a factor of $\gtrsim$\,$3-5$ in TNG-CR as compared to the fiducial run. This also hints towards suppressed outflow rates, as the transport of metal enriched gas from the galaxy and into halo plays an important role in enriching the CGM with metals \citep{peroux2020,pandya2021}. Note however that the metallicity profiles are also offset at the halo centres, albeit to a lower level than the CGM, a direct impact of the tight stellar mass-gas phase metallicity relation \citep{tremonti2004,ma2016,torrey2019}.

Lastly, the insets in all panels show that profiles of individual halos are diverse. A variation of $\lesssim$\,$1$\,dex at fixed distance is seen in most cases, often correlating with the mass of the parent halo. For instance, more massive halos typically have the hottest CGMs, most pressure support and strongest magnetic fields. In cases like metallicity and pressure ratio, the scatter is largely random with respect to halo mass, suggesting that these variations are driven by other factors.

\begin{figure}
    \centering
    \includegraphics[width=9cm]{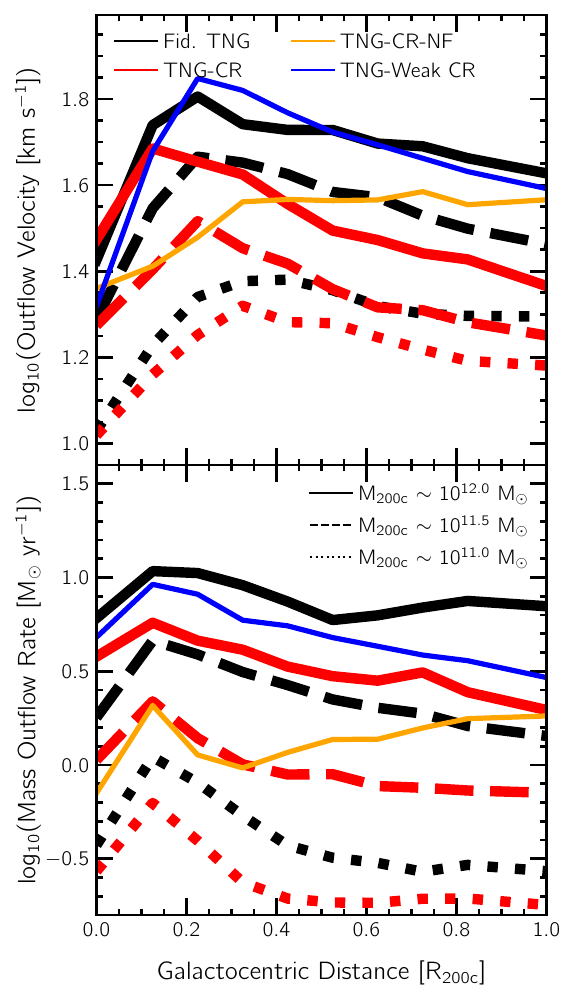}
    \caption{Mass outflow rates (lower panel) and outflow velocity (upper panel) as a function of distance. Colors show different runs, and line styles different halo mass bins. In all mass bins, the outflow rate and velocities are suppressed throughout the halo in TNG-CR with respect to the fiducial TNG case. At the Milky Way mass range, the outflow rate is suppressed further in TNG-CR-NF, and the radial trend reverses, leading to larger velocities and outflow rates in the outer versus inner halo ($\gtrsim$\,$0.5$\,R$_{\rm{200c}}$). TNG-Weak CR is typically intermediate between TNG and TNG-CR. Note, however, that the stellar masses are different in the CR runs as compared to fiducial TNG (see main text).}
    \label{fig:massOutflowRate}
\end{figure}

In Fig.~\ref{fig:massOutflowRate}, we dig deeper into the outflow dynamics of gas. The top panel shows outflow velocity, while the lower shows mass outflow rate. At a galactocentric distance $r$, the latter is defined as $\dot{M}_{\rm{out}}(r) = 1 / \Delta r \sum m_i v_{\rm rad,i}$ where the sum is over all gas cells $i$ with mass $m_i$ and radial velocity $v_{\rm rad,i}$ that have positive radial velocities ($v_{\rm rad,i} > 0$ i.e. outflowing) and reside within a shell of thickness $\Delta r$ at a radius $r$ (i.e. $|r_{\rm i} - r| \leq \Delta r/2)$. The outflow velocity is computed as the mean radial velocity of the same subset of gas cells. We fix $\Delta r$ to be $0.05$\,R$_{\rm{200c}}$, and note that results are qualitatively similar for other choices of $\Delta r$.

As before, we focus on central galaxies of halos in bins of M$_{\rm{200c}}$\,$\sim$\,$10^{11}$, $10^{11.5}$ and $10^{12}$\,M$_\odot$, shown in dotted, dashed and solid curves, respectively. Curves in black correspond to fiducial TNG, red to TNG-CR, orange to TNG-CR-NF and blue to TNG-Weak CR. Although the latter two are only shown for the Milky Way mass bin, we mention that they portray qualitatively similar trends for the other two mass bins as well. For the Fid. TNG case, we once again restrict the analysis to those halos in which the black hole is active in the quasar mode.

In all three mass bins, the mass outflow rates are suppressed in TNG-CR as compared to TNG. The offset is roughly $0.1-0.2$\,dex in the galaxy, but rises to $\sim$\,$0.3-0.5$\,dex as one transitions further out into the halo. The TNG-Weak CR result (blue) lies between the TNG and TNG-CR values and shows a qualitatively similar behaviour, i.e. increasing from the center of the galaxy to a distance of $\sim$\,$0.1$\,R$_{\rm{200c}}$ (typical extent of the gas disk), and dropping at larger distances. The orange curve, however, is qualitatively different, both in shape and normalisation: while the curve also has a peak at $\sim$\,$0.1$\,R$_{\rm{200c}}$, thereafter it rises towards R$_{\rm{200c}}$ and the outer halo.

In the absence of supernovae and black hole feedback, this outflow is purely driven by gradients in CR pressure \citep[e.g.][]{quataert2022,modak2023,matrinalvarez2023}, and is different in nature from outflows driven by other physical mechanisms \citep{simpson2016,thomas2024}. Most importantly, the mass outflow rate is smaller at all distances as compared to the other curves, i.e. at fixed halo mass, outflows driven solely by CR pressure gradients are overall weaker with respect to other feedback processes. It is however important to note that stellar masses (and star formation rates) are not constant across the different physics variations, and mass loading factors portray additional behaviours (see Fig.~\ref{fig:massLoadingFactor}).

Outflow velocities follow similar trends: in all three mass bins, at most distances, velocities are smaller in the TNG-CR run as compared to TNG. Only at the centre of halos are the two similar, a direct result of the TNG model, where the wind velocity at injection is set by the local velocity dispersion of dark matter \citep{pillepich2018}, and is thus primarily set by the halo mass, modulo any local disturbances. Interestingly, outflow velocities in the TNG-CR-NF run (orange) are in fact larger than the red curve at distance $\gtrsim$\,$0.4$\,R$_{\rm{200c}}$, likely as a result of smaller ambient gas densities which provide lower levels of resistance as the outflow propagates outwards. We speculate that this enhanced outflow velocity, which increases towards larger distances, is the primary driver of large scale bubbles seen in Fig.~\ref{fig:fid_vs_cr_image}, which are absent in TNG-CR since velocities are typically smaller and decay towards the halo outskirts.

\begin{figure*}[ht!]
    \centering
    \includegraphics[width=18cm]{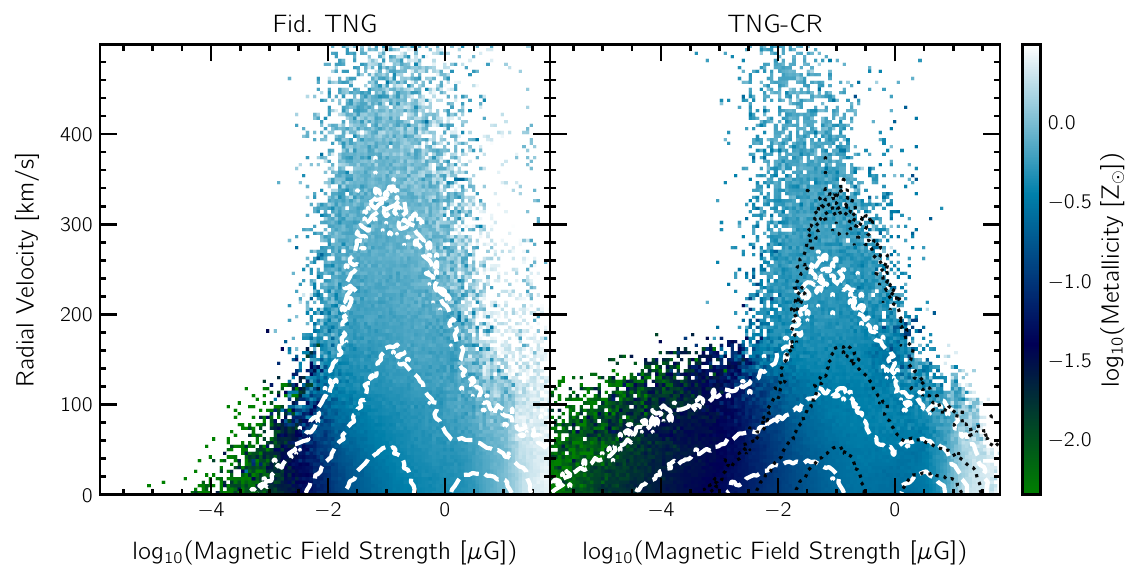}
    \caption{Two dimensional plots of radial velocity (y-axis) as functions of magnetic field strength (x-axis) with background pixels colored by the mean gas phase metallicity, contrasting the Fid. TNG run (left) versus TNG-CR (right). We focus exclusively on outflowing gas (v$_{\rm{rad}}$\,$\geq$\,$0$) within R$_{\rm{200c}}$ of the centrals of Milky Way mass halos, i.e. with satellite gas excised. Dashed white curves show contours drawn at the [1, 10, 50]\,\% levels, increasing in level towards smaller velocities. For ease of visualisation, the dotted black curves on the right overplot the contours from the left panel. A greater fraction of outflows gas in Fid. TNG extends to velocities as large as $O(100)$\,km\,s$^{-1}$, carrying metal enriched gas outwards from the galaxy. The Fid. TNG contours are shifted towards slightly larger field strengths, implying the transport of relatively higher magnetised gas by these outflows.}
    \label{fig:outflow_2d}
\end{figure*}

As outflow rates and velocities in TNG-CR are clearly suppressed with respect to TNG, this will impact magnetic field strengths and metallicites in the CGM. Fig.~\ref{fig:outflow_2d} shows two dimensional plots of radial velocity (y-axis) as functions of magnetic field strength (x-axis) with background color indicating mean gas phase metallicity, contrasting TNG (left) versus TNG-CR (right). We focus exclusively on outflowing gas (v$_{\rm{rad}}$\,$\geq$\,$0$) within R$_{\rm{200c}}$ of the centrals of Milky Way mass halos, i.e. with satellite gas excised. As before, halos in which the black hole is in the kinetic mode are excluded. Dashed white curves show contours drawn at the [1, 10, 50]\,\% levels\footnote{Contours drawn at $x$\% correspond to bins whose relative mass of gas, i.e. normalised by the bin with the maximum mass, is $x$\%.}, increasing in level towards smaller velocities. For comparison purposes, the dotted black curves on the right replicate the contours from the left panel.

Contrasting equi-level contours between the TNG and TNG-CR simulations reveals that a greater fraction of gas in the fiducial TNG model case is outflowing at large velocities, as high as $\sim$\,$O(100)$\,km\,s$^{-1}$. This is in agreement with the upper panel of Fig.~\ref{fig:massOutflowRate}. Background colors further indicate that the slower outflows in TNG-CR carry gas that is less metal enriched, at the level of $0.01 - 0.1 Z_{\rm sun}$. We reiterate the complication, however, that the metallicites of the galaxies themselves differ across the two runs (lower right panel of Fig.~\ref{fig:prof_comp}).

Contrasting contours also shows that outflows in TNG-CR are shifted towards weaker magnetic field strengths. The slower outflows in TNG-CR thus also transport relatively weakly magnetised gas, yielding overall smaller field strengths in the halo (lower left panel of Fig.~\ref{fig:prof_comp}). There is also a clear correlation between weakly magnetised and metal pristine gas, suggesting that the increasing mass outflow rates at larger halocentric distances in the CR models accelerate in-situ halo gas, rather than ejecting ISM gas, resulting in a different mass loading trend with radius \citep[see also][]{mitchell2020}.

\begin{figure*}[ht!]
    \centering
    \includegraphics[width=18cm]{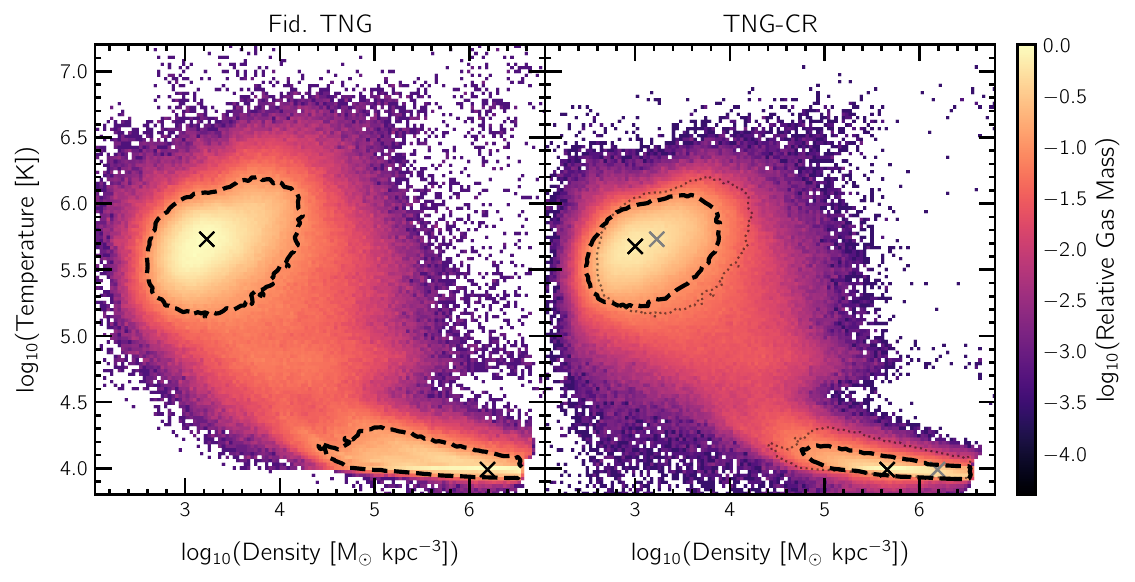}
    \caption{Phase diagrams of gas within R$_{\rm{200c}}$ of Milky Way mass halos in the Fid. TNG run (left) versus TNG-CR (right). Crosses in both panels correspond to (local) maxima of the cold ($<$\,$10^{4.5}$\,K) and hot ($>$\,$10^{5.5}$\,K) phases, while dashed curves show contours drawn at the 10\% level. For ease of visualisation, the dotted curves and gray crosses on the right overplot the corresponding features from the left panel. While the peak density of the cold phase is larger by $\sim$\,$0.3$\,dex in Fid. TNG with respect to TNG-CR, the hot phase peak is only offset by $\sim$\,$0.1$\,dex in density, and $\sim$\,$0.05$\,dex in temperature. Furthermore, contours are more broad in Fid. TNG as compared to TNG-CR, implying larger variation in these gas properties at fixed halo mass.}
    \label{fig:phaseDiagrams}
\end{figure*}

In Fig.~\ref{fig:phaseDiagrams}, we conclude this sub-section with density-temperature phase diagrams of gas within Milky Way-mass halos (M$_{\rm{200c}}$\,$\sim$\,$10^{12}$\,M$_\odot$). The panel on the left corresponds to the fiducial TNG run, and the right to TNG-CR. As before, halos with kinetic-mode i.e. strong SMBHs are excluded. All halos in the mass bin are stacked, and all gas within R$_{\rm{200c}}$ is considered, with satellite gas excised. Color shows the relative mass of gas in each bin, i.e. normalised by the bin with the maximum mass. The black dashed contours are drawn at the 10\% level, and the two crosses correspond to (local) maxima of cold ($<$\,$10^{4.5}$\,K) and hot ($>$\,$10^{5.5}$\,K) gas. For ease of comparison, the dotted curves and gray crosses on the right copy the corresponding features from the left panel.

In density-temperature space, contours enclosing fixed fractions of mass are broader in TNG than in TNG-CR, i.e. the distribution of densities and temperatures is narrower in TNG-CR. Furthermore, there is a systematic shift of the local maxima (i.e. mean values) towards lower densities in TNG-CR. This is more pronounced in the cold phase ($\sim$\,$0.3$\,dex) as opposed to the hot phase ($\sim$\,$0.1$\,dex). While not shown here, we mention that the peak densities of star forming gas are also lower in TNG-CR ($\sim$\,$0.5$\,dex). This provides a compelling explanation for the suppressed star formation rates and stellar masses in the CR models, as lower gas density translates directly into lower SFR  \citep{kennicutt1998}. We attribute the overall lower densities primarily to a direct impact of the added CR pressure support suppressing the amount of gas accretion onto halos and galaxies, resulting in lower densities (Fig.~\ref{fig:massInfowRate}; see also \citealt{buck2020}).

A minor offset in the location of the temperature maxima for the hot phase (of the CGM) is visible -- temperatures are slightly higher in TNG versus TNG-CR ($\sim$\,$0.05$\,dex), in agreement with the top left panel of Fig.~\ref{fig:prof_comp}. This is likely a combination of effects: increased star formation activity in TNG results in more supernovae and thus overall more energy injection from stellar feedback, while the presence of non-vanishing CR pressure at constant total pressure implies somewhat lower thermal pressures. Second order differences are also visible: larger fractions of warm $\sim$\,$10^5$\,K gas, as well as more super-virial $\gtrsim$\,$10^6$\,K gas, in TNG than in TNG-CR.

Overall, while properties like the pressure ratio, magnetic field strength and metallicity are significantly impacted by the inclusion of CRs, either directly or indirectly, the temperature, density and total pressure structure of gas are modified to a smaller extent. However, it is notable that removing both SNe and SMBH feedback yields halos that are under-pressurised and slightly less dense (Fig.~\ref{fig:prof_comp}), and thus some of these results may also be sensitive to the physics employed for other galactic processes. Future projects with additional physics variations will help explore such degeneracies further.

\subsection{Comparison with Other Cosmic Ray Galaxy Simulations}\label{sec:comp_proj}

So far we have qualitatively discussed our results in the context of previous work from the literature. We now contrast our simulations against other projects in a more quantitative manner, in particular with those that employ more complex and sophisticated models for the transport physics of cosmic rays.

We begin with a face-value comparison between the results of our simulations versus those from a set of isolated galaxy simulations presented in \cite{girichidis2024}. In particular, we contrast our findings against their runs which employ a state-of-the-art spectral CR scheme \citep{girichidis2020} coupled to MHD \citep{girichidis2022}. These simulations also include radiative cooling and star formation using the \citealt{springel2003} model, but no strong galactic-scale feedback.

\begin{figure}[ht!]
    \centering
    \includegraphics[width=9cm]{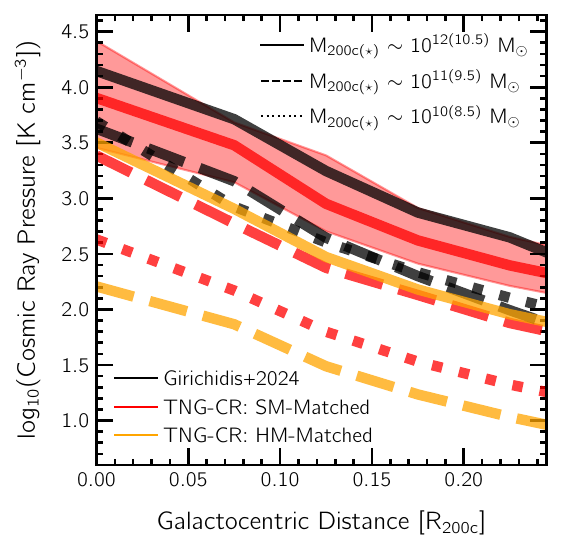}
    \caption{A comparison of spherically-averaged radial profiles of cosmic ray pressure (P$_{\rm{cr}}$\,$=$\,$e_{\rm{cr}}/3$) between TNG-CR and a state of the art spectral CR scheme \protect{\citep{girichidis2024}}. Solid, dashed and dotted curves correspond to M$_{\rm{200c}}$\,$\sim$\,$10^{12}$, $10^{11}$ and $10^{10}$\,M$_\odot$ halos, respectively. In the \protect{\citealt{girichidis2024}} simulations, the galaxies at the centre of these halos have stellar masses M$_{\star}$\,$\sim$\,$10^{10.5}$, $10^{9.5}$ and $10^{8.5}$\,M$_\odot$, shown through the different black curves. Red curves correspond to TNG-CR profiles of galaxies matched with respect to stellar mass, while profiles of the halo-matched TNG-CR sample is shown in orange. Only for the solid red curve, we additionally plot the 16$^{th}$-84$^{th}$ variation across the sample. The stellar mass matched sample, except for the least massive bin, roughly matches the black curves (within a factor of $\sim$\,$1.5$). The halo mass matched sample however is offset by much larger factors, owing to suppressed stellar masses at fixed halo mass.}
    \label{fig:comp_pg}
\end{figure}

Fig.~\ref{fig:comp_pg} shows spherically-averaged radial profiles of cosmic ray pressure (P$_{\rm{cr}}$\,$=$\,$e_{\rm{cr}}/3$) out to a distance of $\sim$\,$0.25$\,R$_{\rm{200c}}$\footnote{We limit the comparison to the inner halo since the \citealt{girichidis2024} simulations are idealized and only evolved for $\sim$\,$3$\,Gyr.}, i.e. focused in on the galaxy, the disk-halo interface, and the inner CGM. Solid, dashed and dotted curves correspond to M$_{\rm{200c}}$\,$\sim$\,$10^{12}$, $10^{11}$ and $10^{10}$\,M$_\odot$ halos, respectively. In the \citealt{girichidis2024} simulations, the galaxies at the centre of these halos have stellar masses M$_{\star}$\,$\sim$\,$10^{10.5}$, $10^{9.5}$ and $10^{8.5}$\,M$_\odot$, shown through the different black curves. Red curves correspond to median TNG-CR profiles of galaxies matched with respect to stellar mass, while profiles of the halo mass matched TNG-CR sample are shown in orange. For the solid red curve alone, we additionally plot the 16$^{th}$-84$^{th}$ variation across the sample.

Starting from a P$_{\rm{cr}}/k_\mathrm{B}$ of $\sim$\,$10^{4.1}$\,K\,cm$^{-3}$ at the centre of the galaxy, the \citealt{girichidis2024} profile for their Milky Way mass halo (solid black) steadily declines by over an order or magnitude to $\sim$\,$10^{2.5}$\,K\,cm$^{-3}$ at $\sim$\,$0.25$\,R$_{\rm{200c}}$. The $M^\star-$matched TNG-CR (solid red) profile shows the same qualitative behaviour, but is vertically offset downwards $\sim$\,0.2\, dex at all distances. At a fixed stellar mass of $10^{10.5}$\,M$_\odot$, the CR pressure support in our TNG-CR runs (with the transport parameters we select) is thus underestimated by a factor of $\sim$\,$1.5$. Note however that the percentile regions are relatively broad, $\sim$\,$\pm$\,$0.4$\,dex at all distances, and the P$_{\rm{cr}}$ profiles in some of these galaxies are in excellent agreement with \citealt{girichidis2024}.

A similar trend is seen between the dashed curves, i.e. for M$_\star$\,$10^{9.5}$\,M$_\odot$ galaxies. At even lower stellar masses (M$_\star$\,$10^{8.5}$\,M$_\odot$; dotted curves), the offset is larger (a factor of $\sim$\,$10$). This difference arises because CR transport at this galaxy stellar mass is dominated by advection for prolonged periods of time in the \citealt{girichidis2024} run, thereby giving rise to greater CR pressure support in the halo. In more massive halos, increased gravity limits advective transport to shorter intervals of time, beyond which diffusion dominates. Assuming a universal constant $v_{\rm{st,eff}}$ at all halo masses and at all cosmic epochs, as is required by the \cite{hopkins2023b} model and adopted in our CR runs, neglects such effects. We note however that P$_{\rm{cr}}$ in TNG-CR at M$_\star$\,$\sim$\,$10^{8.5}$\,M$_\odot$ may also be under-estimated as a result of numerical resolution (see Appendix~\ref{app:res_conv}).

Trends are qualitatively similar with the halo matched sample, although quantitatively different. For instance, the $M_{\rm halo}-$matched (solid orange) is offset low by $\sim$\,$0.6$\,dex at all distances, i.e. by a factor of $\sim$\,$4$ with respect to \citealt{girichidis2024} (solid black), as a result of suppressed stellar masses at fixed halo mass (Fig.~\ref{fig:param_space_var}). The offset is larger for lower halo masses (up to a factor of $\sim$\,$20$), but this is likely due to the stellar mass of the \citealt{girichidis2024} run being larger than the empirically value motivated value (black curve in Fig.~\ref{fig:param_space_var}), due to the idealised nature of their setup. Note that the dotted orange curve, which corresponds to galaxies that are very close to unresolved at our resolution (M$_\star$\,$\lesssim$\,$10^{7.0}$\,M$_\odot$), is absent in this plot.

Overall we conclude that, at fixed stellar mass, our simple CR model and assumed transport parameters result in a reasonable match with the P$_{\rm{cr}}$ profiles of \citealt{girichidis2024}, based on much more sophisticated CR physics. This agreement is reassuring and demonstrates that the net impact of CRs in our runs is plausible, at least from an energetics point of view. As stellar masses are suppressed in TNG-CR, the halo mass matched sample is naturally offset in terms of pressure support.

\begin{figure}[ht!]
    \centering
    \includegraphics[width=9cm]{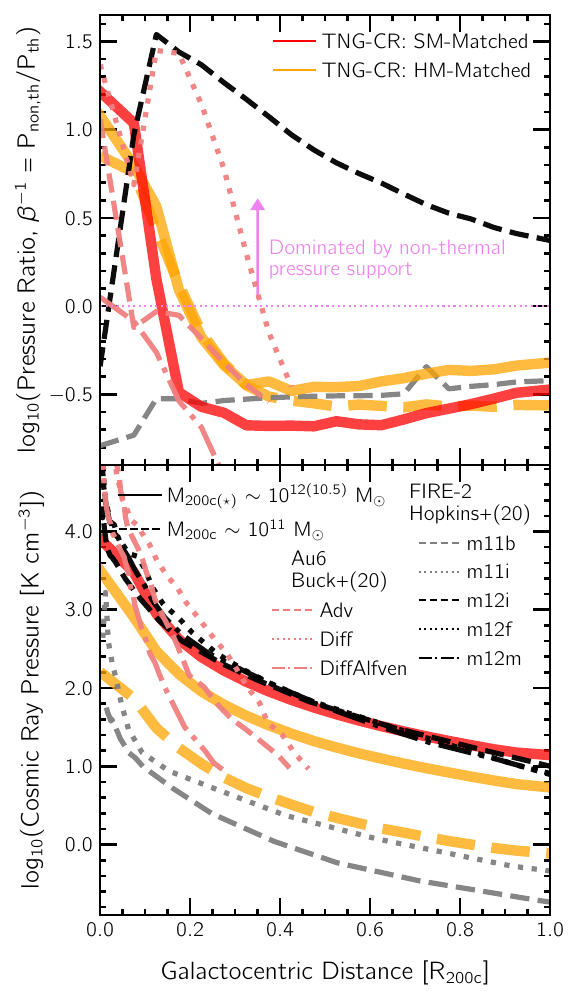}
    \caption{A comparison of spherically-averaged radial profiles of cosmic ray pressure (P$_{\rm{cr}}$; lower panel) and ratio of non thermal-to-thermal pressures (upper panel) between TNG-CR and other cosmological simulations: FIRE-2 and Auriga. Gray and black curves correspond to halos from FIRE-2, coral to multiple CR physics variations of Auriga halo 6 (Au6), and red (stellar mass matched) and orange (halo mass matched) to those from TNG-CR. While the P$_{\rm{cr}}$ profiles are in broad agreement between TNG-CR and FIRE-2, the type of pressure support in these halos is significantly different: in TNG-CR, halos are dominated by non-thermal (thermal) support in the inner (outer) regions, largely independent of halo mass. In contrast, low (high) mass halos in FIRE-2 are almost always dominated by thermal (non-thermal) support. Results of Au6 depend on the CR physics, but profiles of pressure ratios are typically more qualitatively similar to TNG-CR than to FIRE-2.}
    \label{fig:comp_fire}
\end{figure}

In Fig.~\ref{fig:comp_fire}, we contrast our results against those from the FIRE-2 project \citep{hopkins2020,hopkins2023b}. These simulations use a single energy i.e. gray approximation for CRs, they are fully cosmological in nature and hence make it possible to extend the above analysis to larger, CGM scales. The FIRE-2 runs we contrast against employ an explicit CR transport scheme with a constant anisotropic diffusivity $\kappa_\parallel$\,$=$\,$3\,\times\,10^{29}$\,cm$^2$\,s$^{-1}$, and include ideal MHD, supernovae feedback, metal line cooling, amongst other processes \citep[see][]{hopkins2018,hopkins2020}.

These runs are shown through the various gray (m11b, m11i) and black (m12i, m12f, m12m) curves. The former correspond to halos of mass M$_{\rm{200c}}$\,$\lesssim$\,$10^{11}$\,M$_\odot$, while the latter to $\sim$\,$10^{12}$\,M$_\odot$. Profiles of the halo mass matched sample from TNG-CR is shown in orange, and stellar mass matched in red. Note that the latter is shown only for the Milky Way mass, since the less massive FIRE-2 halos have stellar masses M$_\star$\,$\sim$\,$10^8$\,M$_\odot$, at which scale we are limited by numerical resolution (Appendix~\ref{app:res_conv}).

The lower panel focuses on radial profiles of cosmic ray pressure, P$_{\rm{cr}}$. Remarkably, we find that the stellar mass matched sample is in excellent agreement with the corresponding FIRE-2 sample. The halo mass matched sample is offset vertically by $\sim$\,$0.4$\,dex at all distances, a consequence of suppressed stellar masses at fixed halo mass (Fig.~\ref{fig:param_space_var}). The gray curves portray a similar shape as the dashed orange, but are offset vertically below, since these halos are in fact slightly less massive than $10^{11}$\,M$_\odot$. That is, a one-to-one comparison is not possible in terms of halo mass simply due to limited numerical resolution at the low mass end (Appendix~\ref{app:res_conv}). Overall, we conclude that the simple transport model and parameters in TNG-CR give a decent match with expected non-thermal CR pressure support predicted by the FIRE-2 + CRs models.

The upper panel of Fig.~\ref{fig:comp_fire} compares profiles of the ratio of non-thermal to thermal pressure. To guide the eye, the dotted horizontal violet line demarcates the regime of non-thermal pressure dominance (above) from that of thermal support (below). Note that only the m11b and m12i curves from FIRE-2 are available, and so are shown here.

The behaviour of these pressure ratios is markedly different than the total CR pressure profiles. While halos in TNG-CR are dominated by non-thermal pressure support in the inner halo ($\lesssim$\,$0.2$\,R$_{\rm{200c}}$), gas at larger distances is instead supported by thermal pressure, roughly independent of the two halo masses considered here. In stark contrast, the $\sim 10^{12}$\,M$_\odot$ FIRE-2 halo is dominated by thermal support only at the very centre of the galaxy, and transitions into a non-thermal pressure support regime throughout the rest of the halo. The $\sim 10^{11}$\,M$_\odot$ FIRE-2 halo is thermal pressure dominated at all distances, i.e. the halo mass dependence of this finding also differs.

The reason for this inversion of pressure ratio across mass scales in FIRE-2, as well as the difference with respect to TNG-CR, is unclear. As \citealt{ji2020} discuss, the physical properties of gas in the CGM of FIRE-2 galaxies with CRs are substantially altered. Not only does the mean temperature decrease by order of magnitude levels, the trend of temperature with density inverts. Overall, these simulations feature a cooler, CR-dominant CGM, substantially altering predicted observables for e.g. metal ion absorption as well as X-ray emission, and likely representing an extreme case for the impact of CRs which is in tension with such data \citep{ji2020}. 

There are few additional cosmological i.e. non-idealized galaxy formation simulations incorporating CR physics. Notably, \citealt{buck2020} present resimulations of two Auriga Milky Way-mass halos with variations on CR physics. In Fig.~\ref{fig:comp_fire} we also compare to the Au6 run from that work, contrasting simulations assuming pure advection based CR transport (Adv; coral dashed), with added anisotropic diffusion using a diffusion coefficient $\kappa_\parallel$\,$=$\,$10^{28}$\,cm$^2$\,s$^{-1}$ (Diff; coral dotted), and further with the inclusion of Alfv\'en cooling of CRs (DiffAlfven; coral dot dashed). Overall, the CR pressure profiles are steeper than TNG-CR, being larger in the inner halo and lower at $\sim$\,R$_{\rm200c}/2$. With respect to the pressure ratio, the \citealt{buck2020} results depend on the CR physics. Some models yield trends more qualitatively similar to TNG-CR than to FIRE-2, although the Diff case produces a non-thermal pressure dominated inner CGM extending to scales intermediate between the TNG-CR and FIRE-2 result.

In addition to differences in CR physics and CR numerical methods, differences in baryonic feedback models as well as the underlying hydrodynamical scheme, may all give rise to significant differences. The impact of CRs on galaxy and CGM properties clearly eludes current consensus.


\subsection{Discussion and Outlook}\label{sec:outlook}

In TNG-CR, the inclusion of CRs leads to a suppression of stellar masses at fixed halo mass (Fig.~\ref{fig:param_space_var}), a direct consequence of reduced star formation rates over cosmic epochs (Fig.~\ref{fig:prop_comp}). There are at least four possible mechanisms: (1) the added non-thermal CR pressure support may reduce accretion rates of CGM gas onto the galaxy, or the accretion rate of gas into the halo from larger scales \citep{hopkins2020,ruszkowski2023}, (2) the increased ionization and (3) heating terms may reduce the amount and/or density of cold gas in the galaxy or the CGM, or (4) outflows driven by CR pressure gradients \citep{pakmor2016,girichidis2018} may suppress galactic star formation rates. These reflect both preventative and ejective feedback mechanisms.

In our models, options (2) and (3) do not appear to cause a significant impact, as the magnitude of these contributions is subdominant to other relevant ionization/cooling terms. As discussed in Fig.~\ref{fig:massOutflowRate}, our CR driven outflows are not generally as powerful as other feedback-driven outflows. In TNG-CR, we therefore suggest that the added CR pressure support, and the resulting modulation of gas flow rates throughout the galactic baryon cycle, causes the suppression of galactic star formation.

Although it is possible that our results are biased by the limitations of the simple model that we employ, similar qualitative behaviours have been found with more advanced transport schemes \citep{hopkins2020,farcy2022,montero2024}. We refer the reader to \citealt{hopkins2023b} for a discussion on the possible aspects in which this model is likely to fail quantitatively.

As motivated in Section~\ref{sec:param_var}, we have selected transport parameters that reasonably reproduce a subset of realistic galactic-scale properties, while also being in agreement with the narrow range in parameter space proposed by \cite{hopkins2021}. These parameters also offer a decent match to CR pressure support predicted by more advanced models (Section~\ref{sec:comp_proj}). However, certain properties of galaxies in the TNG-CR run, as well as TNG-Weak CR and other variants, are clearly ruled out by available observational constraints. These include the cosmic SFRD, the stellar mass function, and the stellar-to-halo-mass relation (SMHM; Fig.~\ref{fig:prop_comp}). Moreover, the reduced magnetic field strengths and metallicities will alter observable rotation measures (RM) and ionic column density and covering fractions, where observational datasets and constraints are rapidly improving \citep[e.g.][]{wu2024,anderson2024}.

Stellar masses are also suppressed even in the total absence of supernovae and black hole feedback from the TNG model (dotted curve in the right panel of Fig.~\ref{fig:param_space_var}), particularly at the high mass end (M$_{\rm{200c}}$\,$\gtrsim$\,$10^{11}$\,M$_\odot$). This suggests that, independent of feedback physics, the TNG-CR model is too strong in its impact, and furthermore, that no re-calibration i.e. weakening of the TNG feedback model can bring TNG-CR into agreement with observational constraints. 

How can the galaxy properties of TNG-CR be reconciled with observational constraints? For a given set of transport parameters, one approach to reduce the non-thermal CR pressure support is to adopt a lower value for the fraction of CR energy per SNe ($\epsilon_{\rm cr}^{\rm SNe}$). While $\epsilon_{\rm cr}^{\rm SNe}$ was set to $10\%$ for all the runs presented in the main text, Appendix~\ref{app:cr_epsl} explores the impact of varying this fraction. Considering a value of, e.g., $\epsilon_{\rm cr}^{\rm SNe}$\,$=$\,$1$\,\% returns a SMHM relation much closer to emperical constraints. While \citealt{murase2019} infer a constraint of $\epsilon_{\rm cr}^{\rm SNe}$\,$\lesssim$\,$5$\,$-$\,$10$\,\%, this is based on observations of a single SNe, and is also model dependent. More observations of this kind are needed to better constrain $\epsilon_{\rm cr}^{\rm SNe}$ and its possible dependencies.

Within the realm of the \citealt{hopkins2023b} model, an alternate approach may be to relax some subset of the underlying assumptions. In particular, by requiring that CR transport parameters are fixed in space and time, one is likely to miss CR dynamics that depend on halo mass. For instance, \citealt{girichidis2024} suggest that CR transport in low-mass halos (M$_{\rm{200c}}$\,$\sim$\,$10^{10-11}$\,M$_\odot$) is dominated by advection, at least temporarily, while more massive halos are instead dominated by diffusion. This would correspond to a time- (and space-) dependent $v_{\rm{st,eff}}$ that increases (locally) during an advection dominated period, and reduces to smaller values when diffusion is instead the primary transport mechanism. More generally, the transport mode (diffusion/streaming) and the effective transport speed need to be computed based on local hydrodynamic properties, which naturally lead to locally (strongly) varying quantities \citep[see, e.g.][]{jiang2018,thomas2019,thomas2022}. Exploring such a model remains the topic of future work.

Despite the computational expensive, more sophisticated i.e. explicit CR models must also be explored in large-volume cosmological simulations. This includes more realistic transport, and spectral schemes that can capture the energy-dependent effects of the CR population. It remains to be seen if, and how, realistic galaxy populations can be obtained with explicit and physically plausible CR scenarios.


\section{Summary and Conclusions}\label{sec:conc}

In this work, we have used a suite of (25\,Mpc\,h$^{-1}$)$^3$ cosmological magnetohydrodynamical boxes to explore the impact of cosmic rays (CRs) on the evolution of galaxies. To do so, we couple the IllustrisTNG galaxy formation model \citep{weinberger2017,pillepich2018} with a simple sub-grid model for CR transport \citep{hopkins2023b}. Our main findings are as follows:

\begin{enumerate}
    \item We contrast a fiducial TNG simulation with multiple models including CRs: a fiducial but strong TNG-CR parameterization, a TNG-Weak CR alternative, and CRs but no other feedback control case. In general, the quantitative impact of CRs on galaxy properties depends on the values of our two CR transport parameters -- the effective diffusion coefficient and streaming velocity.
    
    \item The inclusion of CRs reduces star formation rates, suppressing galaxy stellar masses at fixed halo mass. This effect can be as large as a factor of $\sim$\,$10$ at the Milky Way mass range, decreasing to $\sim$\,2 for $M^\star \sim 10^9$\,M$_\odot$ dwarf galaxies (Fig.~\ref{fig:param_space_var}).

    \item Our fiducial TNG-CR model brings several summary statistics and scaling relations into strong tension with observations. These include the star formation rate density (SFRD), galaxy stellar mass function (SMF), and stellar mass to halo mass relation. Others, such as the galaxy size-mass relation and halo-scale gas fractions are not qualitatively changed, and remain in broad agreement with data. The growth of supermassive black holes is reduced, due to lower densities in the central star-forming interstellar medium, flattening black hole mass relations (Fig.~\ref{fig:prop_comp}).

    \item The inclusion of CRs does not significantly impact the temperature, density or (total) pressure structure of the circumgalactic medium, as compared to the TNG model. Halo gas, however, is supported by non-thermal pressure to a larger extent, and shows lower levels of magnetisation and metal enrichment with CRs (Figs.~\ref{fig:fid_vs_cr_image}, \ref{fig:fid_vs_cr_vs_nf_halo_image} and \ref{fig:prof_comp}).

    \item The primary physical driver of the differences between TNG and TNG-CR is that: (i) CRs add appreciable non-thermal pressure support to galactic halos. This suppresses gas inflow rates into both the halo and galaxy. Simultaneously, (ii) CRs reduce the mass outflow rates and velocities of feedback-driven winds (Figs.~\ref{fig:massOutflowRate} and \ref{fig:outflow_2d}).

    \item An appreciable CR energy density permeates galaxies, galactic halos, and extends into the intergalactic medium. We identify a variety of cosmic ray driven, and non-thermal pressure supported, outflows and bubble-like features at a variety of scales, from $~10-50$\,kpc galactic outflows to $\sim$\,Mpc size halo-scale structures (Figs.~\ref{fig:mainVisImage}, \ref{fig:fid_vs_cr_image}, and ~\ref{fig:fid_vs_cr_vs_nf_halo_image}).

    \item The phase structure of gas within the virial radii of Milky Way-like halos is qualitatively unchanged. The presence of CRs lowers the mean density (by $\sim$\,$0.1$\,dex) and temperature ($\sim$\,$0.05$\,dex) of the CGM (Fig.~\ref{fig:phaseDiagrams}).

    \item The assumed transport parameters offer a good match with the cosmic ray pressure profiles predicted by more advanced CR models -- both of isolated galaxy setups with a spectral scheme \citep{girichidis2024}, and the FIRE-2 cosmological simulations. However, the ratio of non-thermal to thermal pressure, and overall impact on the physical properties of the CGM, is starkly different in our simulations as compared to FIRE-2 (Figs.~\ref{fig:comp_pg} and \ref{fig:comp_fire}).
\end{enumerate}

This work is an initial study on the impact cosmic rays have across the galaxy population. It represents our first attempt at \mbox{``illuminating the IllustrisTNG universe with cosmic rays''}, and is a first step towards including CR physics in large volume cosmological simulations, an important future direction for numerical models of galaxy formation and evolution. A key and unresolved caveat is that the galaxy population of our TNG-CR model is in significantly more tension with observations than the original TNG model without CRs. Future work will advance our current modeling by incorporating more advanced treatments of cosmic ray production and transport physics.

\section*{Data Availability}

The original IllustrisTNG simulations are publicly available and accessible at \url{www.tng-project.org/data} \citep{nelson2019b}, where the variations presented here including cosmic ray physics will be made public in the future. Data directly related to this publication is available on request from the corresponding authors. This work has benefitted from the \texttt{scida} analysis library \citep{byrohl2024}.

\begin{acknowledgements}
We thank Annalisa Pillepich for insightful discussions and sharing relevant data, and Matthew C. Smith for valuable inputs on the tree walk implementation. RR and DN acknowledge funding from the Deutsche Forschungsgemeinschaft (DFG) through an Emmy Noether Research Group (grant number NE 2441/1-1). RR is a Fellow of the International Max Planck Research School for Astronomy and Cosmic Physics at the University of Heidelberg (IMPRS-HD). PG acknowledges funding by the European Research Council via the ERC Synergy Grant ``ECOGAL'' (project ID 855130). This work has made use of the VERA supercomputer of the Max Planck Institute for Astronomy (MPIA) operated by the Max Planck Computational Data Facility (MPCDF), and of NASA's Astrophysics Data System Bibliographic Services. 
\end{acknowledgements}

\bibliographystyle{aa}
\bibliography{references}


\appendix
\section{Impact of Alfv\'en Heating}\label{app:alfven_heating}

CRs that propagate faster than the Alfv\'en-velocity excite Alfv\'en-waves through the CR streaming instability \citep{kulsrud1969}, effectively heating up gas as these waves dampen on short-time scales \citep{wiener2017,ruszkowski2017}. For the approximations made in the derivation of the CR sub-grid model, the corresponding heating term is given by $\Lambda_{\rm{alf}}$\,$=$\,$|\vec{v_{\rm{alf}}} \dcdot \vec{\nabla \rm{P_{cr}}}|$/3 \citep{hopkins2023b}, where $|\vec{v_{\rm{alf}}}|$\,$=$\,$|\vec{B}|/(4 \pi \rho)^{1/2}$ is the magnitude of the Alfv\'en-velocity.

\begin{figure}
    \centering
    \includegraphics[width=9cm]{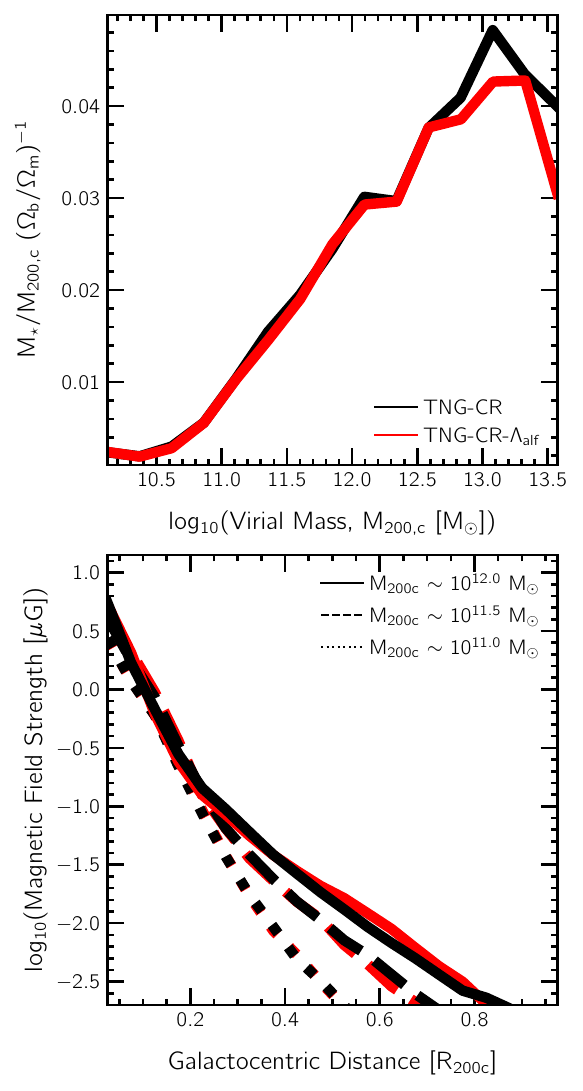}
    \caption{The impact of including Alfv\'en heating. The top panel shows the stellar to halo mass relation, while the lower panel shows radial profiles of magnetic field strength. Colors correspond to different variation runs, and line styles in the to lower panel to different mass bins. Minimal difference is seen between the two runs, suggesting that the inclusion of Alfv\'en heating has a minimal impact under the framework of this simple model.}
    \label{fig:app_alf_cool}
\end{figure}

As mentioned in Section~\ref{ssec:methods_model}, this process was not included in the results presented in the main body of the paper. In Fig.~\ref{fig:app_alf_cool}, we briefly explore the impact of adding this heating term. The upper panel shows the stellar to halo mass relation, while radial profiles of magnetic field strength for three different halo mass bins are shown in the bottom. Colors correspond to different variation runs, and line styles in the to lower panel to different mass bins.

In terms of both stellar mass and magnetic field strength there is minimal difference between the two runs, except for the high halo mass end in the former (M$_{\rm{200c}}$\,$\gtrsim$\,$10^{13}$\,M$_\odot$). This is also qualitatively the case for all other properties considered in this work (not shown). This suggests that the inclusion of Alfv\'en heating has a largely subdominant impact under the framework of the model we employ. We speculate that this may be due to the relative simplicity of the model, i.e. the computation of $\rm{P_{cr}}$ through a distance-modulated locally-attenuated approach results in typically small values of $\vec{\nabla \rm{P_{cr}}}$.

However, even simulations run with more sophisticated CR transport models suggest that the Alfv\'en cooling time is typically larger than its hadronic counterpart in most regions except the very centre of galaxies \citep{girichidis2024}. This would imply that the effect of this added heating term is most pronounced on $\lesssim$\,kpc scales. This regime is both poorly resolved and modulated by our sub-grid ISM pressurization model. It is also where the \citealt{hopkins2023b} CR model is severely limited. More sophisticated galaxy formation, ISM, and CR models can revisit this question.

\section{Mass Loading Factors and Inflow Rates}\label{app:mass_flow}

As discussed in Fig.~\ref{fig:massOutflowRate}, at fixed halo mass, gas outflow rates are suppressed in TNG-CR as compared to the fiducial TNG run. However, stellar masses (and star formation rates) vary across the different physics variations, meaning that the amount of energy available for stellar feedback driven outflows varies substantially between simulations. This makes interpretation non-trivial.

\begin{figure}
    \centering
    \includegraphics[width=9cm]{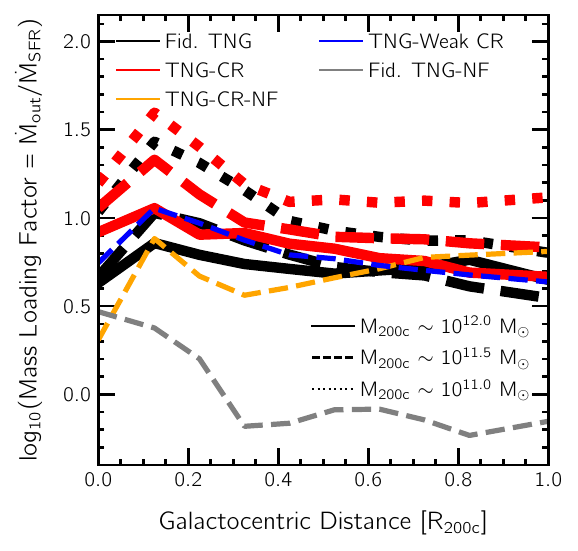}
    \caption{Similar to the bottom panel of Fig.~\ref{fig:massOutflowRate}, but normalised by the star formation rate of the galaxy, i.e. the mass loading factor. In all halo mass bins, values are larger in TNG-CR as compared to Fid. TNG. The added outflow driven by the presence of CRs thus boosts the mass loading factor, albeit with smaller mass outflow rates since star formation rates as diminished.}
    \label{fig:massLoadingFactor}
\end{figure}

To better unravel this effect, Fig.~\ref{fig:massLoadingFactor} instead shows the mass loading factor, i.e. the mass outflow rate normalised by the star formation rate of the galaxy. Halos in bins of mass M$_{\rm{200c}}$\,$\sim$\,[10$^{11.0}$, 10$^{11.5}$, 10$^{12.0}$]\,M$_\odot$ are shown through dotted, dashed and solid curves, respectively. Results from TNG are shown in black, and TNG-CR in red. Only for the M$_{\rm{200c}}$\,$\sim$\,10$^{11.5}$\,M$_\odot$ bin, we additionally show curves from TNG-CR-NF (orange) and TNG-Weak CR (blue), and also from a run which uses the TNG model but with no feedback, i.e. no stellar and black hole feedback, as well as no CRs (Fid. TNG-NF; gray).

Contrasting the various black versus red curves, higher mass loading factors are seen in the TNG-CR runs, in all three halo mass bins. Outflows driven by the CR pressure gradients, added to the already existing SNe and black hole driven feedback in TNG, thus produce higher mass loading factors. Due to smaller stellar masses and star formation rates, this however produces a smaller outflow rate (Fig.~\ref{fig:massOutflowRate}).

In the TNG-CR-NF run, mass loading factors are smaller in the inner halo ($\lesssim$\,$0.5$\,R$_{\rm{200c}}$) as compared to TNG and TNG-CR, and values are comparable to the TNG-CR run only at the very outskirts of the halo. With our simple CR model and assumed transport parameters, outflows driven solely by CR pressure gradients are typically weaker than other feedback processes in TNG \citep[see also][]{hanasz2013,dashyan2020}. Similar to our general finding in the main text, the TNG-Weak CR case (blue curve) is intermediate of between TNG and TNG-CR. Lastly, mass loading factors in the `Fid. TNG-NF' run (no feedback of any kind; gray) are by far the smallest, as expected.

\begin{figure}
    \centering
    \includegraphics[width=9.2cm]{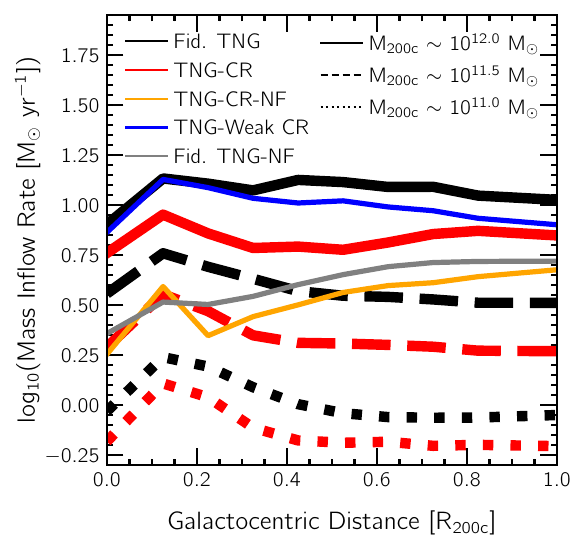}
    \caption{Similar to the bottom panel of Fig.~\ref{fig:massOutflowRate}, but the inflow rate of gas. In all halo mass bins, and at all galactocentric distances, inflow rates are smaller in TNG-CR as compared to Fid. TNG, signifying lower levels of gas infall through the halo and into the galaxy.}
    \label{fig:massInfowRate}
\end{figure}

As a final point of discussion on gas flows, Fig.~\ref{fig:massInfowRate} measures inflow rates of gas through the halo. Analogous to our measurement of mass outflow rates, at a galactocentric distance $r$, the inflow rate is defined as
\begin{equation}
    \dot{M}_{\rm{in}}(r) = \frac{1}{\Delta r} \sum\limits_{\substack{i=0 \\ |r_i - r| \leq \Delta r / 2 \\ v_{\rm{rad},i} < 0} }^n - m_i \times v_{\rm{rad},i}
\label{eq:inflowRate}    
\end{equation}
where notations carry the same definitions as earlier. As with Fig.~\ref{fig:massLoadingFactor}, halos in bins of mass M$_{\rm{200c}}$\,$\sim$\,[10$^{11.0}$, 10$^{11.5}$, 10$^{12.0}$]\,M$_\odot$ are shown through dotted, dashed and solid curves, respectively. Results from TNG are shown in black, and TNG-CR in red. Only for the M$_{\rm{200c}}$\,$\sim$\,10$^{12.0}$\,M$_\odot$ bin, we additionally show curves from TNG-CR-NF (orange) and TNG-Weak CR (blue), and also from Fid. TNG-NF (gray).

Similar to Fig.~\ref{fig:massOutflowRate}, in all halo mass bins and at all distances, inflow rates are smaller in TNG-CR as compared to TNG. On one hand, this is linked to the reduced outflow rates, that then result in lower levels of circulation and fountain flows \citep{oppenheimer2008,peroux2020b}. However, contrasting the gray (no feedback whatsoever) versus orange (only CRs, no SNe or AGN feedback) curves suggests that the added non-thermal CR pressure support may be an important driver of this trend, i.e. even in the total absence of SNe, black hole or CR driven outflows, and thus a complete lack of gas circulation between the galaxy and the CGM, the Fid. TNG-NF simulation produces larger inflow rates than the TNG-CR-NF case. Similar to discussions in the main text, this suppression of gas inflow is due to the added CR non-thermal pressure support.

\section{Assessing Convergence of Numerical Resolution}\label{app:res_conv}

As mentioned in Section~\ref{sec:methods}, all results presented in the main text are based on the L25n512 suite of simulations, i.e. those run at an average baryonic mass resolution of $\sim$\,$10^{6}$\,M$_\odot$ (equivalent to TNG100-1). We here assess numerical convergence by running analogous TNG-CR boxes at increasingly coarser resolutions.

\begin{figure}
    \centering
    \includegraphics[width=9cm]{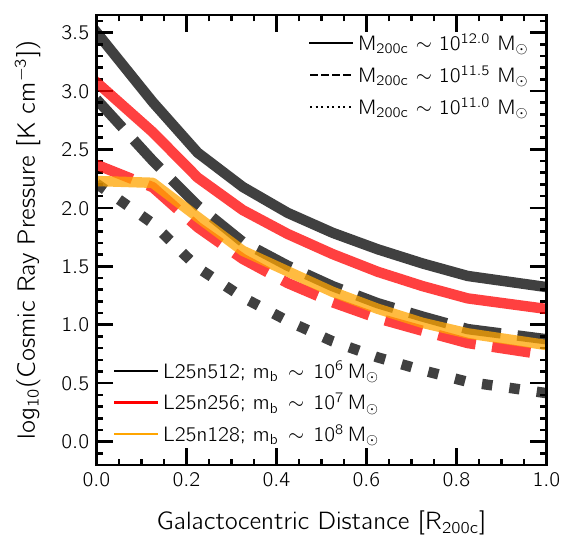}
    \caption{Comparison of CR pressure (P$_{\rm{cr}}$) profiles across simulations run at varying numerical resolutions. For a given halo mass bin, values of P$_{\rm{cr}}$ are typically smaller at coarser resolution, a direct impact of suppressed star formation rates. Note that a subset of curves are absent, as these produce profiles of P$_{\rm{cr}}$\,$\sim$\,$0$ as a result of limited resolution.}
    \label{fig:res_conv}
\end{figure}

In Fig.~\ref{fig:res_conv}, we show CR pressure (P$_{\rm{cr}}$) profiles of central galaxies from three different mass bins: M$_{\rm{200c}}$\,$\sim$\,[10$^{11.0}$, 10$^{11.5}$, 10$^{12.0}$]\,M$_\odot$ in dotted, dashed and solid curves, respectively. Results from L25n512 are shown in black, while lower resolution counterparts L25n256 (m$_{\rm{b}}$\,$\sim$\,$10^{7}$\,M$_\odot$) and L25n128 (m$_{\rm{b}}$\,$\sim$\,$10^{8}$\,M$_\odot$) in red and orange, respectively. Note that a subset of curves are absent, as these produce profiles of P$_{\rm{cr}}$\,$\sim$\,$0$ as a result of limited resolution. 

At a given halo mass, values of P$_{\rm{cr}}$ are typically smaller at coarser resolution. For instance, the red solid curve is offset by $\sim$\,$0.15$\,dex with respect to the black solid curve throughout most of the halo, with a slightly larger difference of $\sim$\,$0.4$\,dex at the very centre of the galaxy. We understand this to be a direct impact of lower star formation rates at coarser levels of resolution \citep{pillepich2018}.

A similar trend is seen while contrasting black and red dotted curves, i.e. the M$_{\rm{200c}}$\,$\sim$\,10$^{11.5}$\,M$_\odot$ bin. At an even lower halo mass of M$_{\rm{200c}}$\,$\sim$\,10$^{11.0}$\,M$_\odot$, the red (dotted) curve is absent, implying that these halos are the smallest ones that one could explore in the L25n512 run, which is the cutoff that we use throughout the main body of this work (Section~\ref{sec:methods}).

\section{Impact of varying $\epsilon_{\rm cr}^{\rm SNe}$}\label{app:cr_epsl}

Results presented throughout the main text are based on simulations where $\epsilon_{\rm cr}^{\rm SNe}$, the CR energy fraction available from SNe, is set to $10$\,\%. We here explore the impact of varying this fraction.

\begin{figure}
    \centering
    \includegraphics[width=9cm]{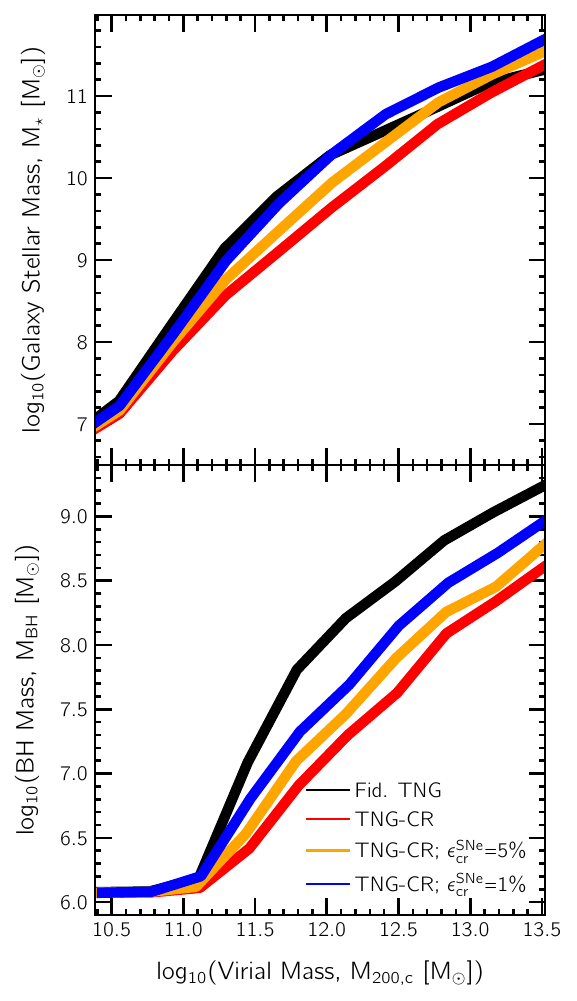}
    \caption{A comparison of the stellar mass-(upper panel) and black hole mass-(lower panel) halo mass relation between simulations with varying values of $\epsilon_{\rm cr}^{\rm SNe}$. Lower values effectively result in weaker CR pressure support in the halo, yielding a more realistic SMHM relation.  However, the growth of SMBHs is still suppressed in all cases, due to high CR pressures in the very center regions of star-forming galaxies.}
    \label{fig:epsl}
\end{figure}

In Fig.~\ref{fig:epsl}, we compare the ($z$\,$=$\,$0$) stellar mass-(upper panel) and black hole mass-(lower panel) halo mass relations from several simulations. The black curve shows results from the TNG run, while the TNG-CR is shown in red. Orange and blue correspond to new runs with $\epsilon_{\rm cr}^{\rm SNe}$\,$=$\,$5$ and $1$\,\%, respectively.

Reducing the value of $\epsilon_{\rm cr}^{\rm SNe}$ yields a SMHM relation closer to the black curve, i.e. to empirical constraints. For instance, the $\epsilon_{\rm cr}^{\rm SNe}$\,$=$\,$1$\,$\%$ curve is in excellent agreement with the black curve up to a halo mass of $10^{12}$\,M$_\odot$. We understand this to be a result of lower amounts of CR pressure support in the halo (not shown), thus allowing a greater amount of gas to accrete on to galaxies, yielding larger star formation rates.

However, the most massive halos above $10^{12}$\,M$_\odot$ have even larger stellar masses than TNG for the lower energy fraction cases (blue and orange). This is a result of diminished black hole masses (lower panel), thus delaying the onset of kinetic mode BH feedback, that is the main quenching mechanism in massive halos in TNG. While CR pressure support in the halo is reduced, the added non-thermal component at the centre of galaxies is still large enough that SMBH accretion rates \citep[via a Bondi model;][]{weinberger2017} are appreciably reduced, thereby limiting their growth. 

While \citealt{murase2019} propose a constraint of $\epsilon_{\rm cr}^{\rm SNe}$\,$\lesssim$\,$5$\,$-$\,$10$\,\%, this is based on observations of a single SNe, and also model dependent. More observations of this kind are needed to assess whether the $\epsilon_{\rm cr}^{\rm SNe}$\,$=$\,$1$\,\% case is indeed realistic. In such a model, one would additionally need to adjust the models for SMBH growth and feedback in order to reproduce TNG galaxy properties at the high-mass end.

\end{document}